\documentclass[12pt]{article}

\usepackage{amsmath, amsthm, amssymb}
\usepackage{graphicx}
\usepackage{color}

\newtheorem{example}{Example}

\title{The relaxation of a family of broken bond crystal surface models}

\author{Jeremy L. Marzuola\footnotemark[1]~  and Jonathan Weare\footnotemark[2]}
\begin{document}

\footnotetext[1]{Department of Mathematics, UNC-Chapel Hill}
\footnotetext[2]{Department of Mathematics, University of Chicago}

\maketitle

\begin{abstract}
We study the continuum limit of a family of kinetic Monte Carlo models of crystal surface relaxation that includes both the solid-on-solid and discrete Gaussian models.  With computational experiments and theoretical arguments we are able to derive several partial differential equation limits identified (or nearly identified) in previous studies and to clarify the correct choice of surface tension appearing in the PDE and the correct scaling regime giving rise to each PDE.  We also provide preliminary computational investigations of a number of interesting qualitative features of the large scale behavior of the models.
\end{abstract}

\section{Introduction}
\label{sec:intro}

Characterizing the evolution of a crystal surface is a worthy goal given the importance of crystal films in many modern electronic devices (e.g. mobile phone antennae).  In this paper we explore the evolution of a family of very simple atomistic models of crystal evolution in certain macroscopic scaling limits.  The family of atomistic models includes the well known solid-on-solid (SOS) model \cite{Binh,pimpinelli1999physics} and is remarkable, given its simplicity, for its close relation to models in widespread use in large scale simulations of crystal evolution  (see e.g. \cite{simbib1,simbib2,simbib3,simbib4,simbib5,simbib6,simbib7} for recent studies).  

While the large scale and qualitative properties of mesoscopic, ordinary differential equation (ODE) models for terraced crystal surfaces, have been studied by many authors (see e.g. \cite{AlHajjShehadehKohnWeare:2011, EYip:2001:contepitaxy,FokRosalesMargetis:2008,IsraeliJeongKandelWeeks:2000:stepscaling,MargetisNakamura:2011} and the references therein), similar investigations of microscopic, kinetic Monte Carlo (KMC) models seem less common.  A notable exception is the paper by Krug, Dobbs, and Majaniemi, \cite{KDM}, on the continuum (large crystal) limit of the SOS model in 1+1 dimensions.  The present work is motivated
by that study.  For a study of the relationship between KMC models and ODE models of terraced surfaces see \cite{PatroneMargetis:2013:KMCODE}.

The authors of \cite{KDM} give informal arguments suggesting a partial differential equation (PDE) governing the evolution of the SOS model in the continuum limit.  We provide a different informal (if slightly less so) argument justifying the same limiting equation as well as provide more extensive numerical supporting evidence.  
Arguments in the last section of \cite{KDM} actually suggest an alternative, and very different, PDE limit for the SOS model.  This PDE has an unusual exponential nonlinearity.  We show that a PDE with a very similar (but not the same) exponential non-linearity can be derived in a particular, non-standard, macroscopic scaling limit.  Our informal argument in this scaling regime is similar to the argument in the standard regime and is again bolstered by numerical simulations. The two PDE are roughly consistent in an appropriate asymptotic sense.

In addition to the two PDE identified in \cite{KDM}, Haselwandter and Vvedensky, in \cite{vvedensky}, suggest another PDE for the macroscopic dynamics, albeit in a slightly different limit.  The goal of this paper is to, through a careful numerical and theoretical investigation, clearly identify the correct PDE limits and how they arise in different limiting regimes.

The paper is organized as follows.  In Sections \ref{sec:background} and \ref{sec:micro}, we describe in detail the family of atomistic models that we consider.  In Section \ref{sec:PDElimits} we  present the relevant PDE limits along with their similarities and differences to results in the literature.  In Section \ref{s:numerics} we give numerical evidence supporting our claims.  Lastly, we offer our (informal) derivation of the PDE limits in Section \ref{sec:PDEder}.

\subsection{Acknowledgments}  
The authors wish to thank Robert V. Kohn, Dionisios Margetis, Peter Smereka, and Herbert Spohn for helpful discussions throughout the preparation of this work as well as Sandeep Sarangi from UNC Computing for assistance with running simulations on the UNC {\it Killdevil} cluster.  Peter Smereka brought reference \cite{KDM} to our attention and was particularly instrumental in the formulation of this project.  J.L.M. was partially supported by an NSF Postdoc and an IBM Junior Faculty development award.  He in addition acknowledges the Courant Institute and the University of Chicago for their gracious hosting during parts of this work.  JW was supported by the NSF through award DMS-1109731.

\section{Background}
\label{sec:background}

The evolution of a crystal is most naturally (and most accurately) captured by ab-initio molecular simulation, i.e. by resolving the fluctuations and bond breaking/formation events of the entire crystal.  Unfortunately such simulations are not practical at large scales.  If we imagine that the evolution of the crystal surface proceeds by rare (on the time scale of atomistic fluctuations) ``hopping'' events in which an atom breaks the bonds with its neighbors and moves from one position on a crystal lattice to a nearby position then it is reasonable to attempt to resolve only the presence or absence of an atom at each lattice position.  The family of microscopic models that we consider here takes this one step further, only describing the evolution of the surface of the crystal and ignoring important features such as vacancies, dislocations, and substrate interaction.  

Despite their deficiencies, 
versions of these so-called broken-bond models have found widespread use in large scale simulation and, as we will see, their relative simplicity makes them amenable to analysis.  In \cite{KDM} the authors considered the macroscopic evolution of a model nearly identical to the one we will soon describe in detail.  That paper serves as the motivation for the current work.  The authors of \cite{KDM} suggest that, appropriately rescaled, the evolution of the surface height of a large crystal in $1+1$ dimensions (one spatial and one time dimension)  can be described by the partial differential equation
\begin{equation}
\label{eqn:krugpde1}
\partial_t h = -K \frac12 \partial_x^3 \left[ \sigma(\partial_x h) \right],
\end{equation}
where $K$ is an inverse temperature and $\sigma(u)$ is a free energy of the surface slope that will be defined precisely later.
While they provide a direct informal argument to justify this conclusion, arguments at the end of \cite{KDM} also suggest that the PDE 
\begin{equation}
\label{eqn:krugpde2}
\partial_t h = \frac12 \partial_x^2\, e^{ -K \partial_x \left[ \sigma(\partial_x h)\right]},
\end{equation}
describes the surface evolution at large scales.  As the authors of \cite{KDM} point out, equation \eqref{eqn:krugpde1} is the small curvature
limit of equation \eqref{eqn:krugpde2}.

In \cite{vvedensky} the authors derive yet another  PDE limit.  That PDE has the form in \eqref{eqn:krugpde1} but differs from 
the result in \cite{KDM} in the definition of the surface free energy term $\sigma.$  The difference is the result of an additional approximation in 
\cite{vvedensky}.  Those authors first consider the limiting behavior of the lattice model as the lattice constant becomes small and time is scaled accordingly.  The resulting approximate microscopic model is an over-damped Langevin diffusion
 for continuous valued height variables at each lattice site.  The large lattice limit of such models have been studied extensively by Funaki and co-workers in, for example, \cite{funakispohn} and, in the appropriate scaling, yields the PDE limit reported in \cite{vvedensky}.

This paper provides arguments and numerical evidence confirming \eqref{eqn:krugpde1} as the correct large scale limit.  We also provide arguments and numerical evidence establishing a PDE similar to \eqref{eqn:krugpde2} (the PDEs differ in the definition of $\sigma$) as the correct large scale limit in an alternative scaling corresponding to large crystals with very rough surfaces.  But before we state our conclusions more precisely we need to describe the family of microscopic models in detail.

\section{The microscopic model}
\label{sec:micro}

We will view the crystal surface as a function  $h_N(t,\alpha)$  of time $t\in [0,\infty]$ and position on the periodic lattice $\alpha\in \mathbb{T}^d_N = \left( \mathbb{Z}/ N\mathbb{Z}\right)^d,$
 with  values in $\mathbb{Z}.$  The symbol $h_N(\alpha)$ without the $t$ argument will occasionally be used to refer to a generic crystal surface.
Let $V: \mathbb{Z} \rightarrow \mathbb{R}$ be a non-negative, strictly convex, symmetric function.  The most common choice in the literature on the physics of crystal surfaces is $V(z) = |z|,$ which is referred to as the solid-on-solid or SOS model.  Other choices of $V$ have been studied as well.  For example, features of the discrete Gaussian model, $V(z)=z^2,$ were examined in \cite{ChuiWeeks:1976:discreteGaussianmodel}.

Define the 
vectors $e_i$ by
\[
(e_i)_j = \begin{cases}
1, & j=i\\
0, & j\neq i
\end{cases}
\]
 and for any function $g:\mathbb{T}^d_N\rightarrow \mathbb{R}$ define 
 the symbols $\nabla_i^+ g (\alpha)$ and $\nabla_i^-g (\alpha)$ by
\[
\nabla_i^+ g (\alpha) =  g(\alpha+e_i) - g(\alpha)\qquad\text{and}\qquad \nabla_i^- g(\alpha) = g(\alpha)-g(\alpha-e_i).
\] 
  The equilibrium probability for the surface gradients $\nabla_i^+ h_N(\cdot)$  is
\[
\rho_N\left(  \nabla^+ h_N(\cdot) \right) \propto \exp\left({- K \sum_{\substack{\alpha \in \mathbb{T}^d_N\\ i\leq d}} V(\nabla_i^+ h_N(\alpha))}\right).
\]
Note that our assumption that $V$ is symmetric obviates inclusion of terms in the sum involving $\nabla_i^-h_N(\cdot).$

 The corresponding equilibrium probability measure for the actual surface is not well defined without constraining some additional feature of the surface such as its average height (the total mass of the crystal).  Here we will be interested in the dynamics of crystal surfaces for which the total mass
\[
m = \sum_{\alpha \in \mathbb{T}^d_N} h_N(\alpha)
\]
remains constant.  Restricting our attention to these surfaces we define the equilibrium measure,
\[
\rho_N^m \left( h_N\right) \propto \begin{cases} \exp\left( -  K \sum_{\substack{\alpha \in \mathbb{T}^d_N\\ i\leq d}} V(\nabla^+_i h_N(\alpha))\right) & \text{if } \sum_{\alpha \in \mathbb{T}^d_N}h_N(\alpha)= m\\
0 & \text{otherwise}.
\end{cases}
\]

Our dynamics will be specified by a continuous time Markov jump process.  The process evolves by jumps of the form
\[
h_N\mapsto J_\alpha^\beta h_N,
\]
where
\[
J_\alpha^\beta = J_\alpha J^\beta
\]
with
\[
J_\alpha h_N(\gamma)  =
\begin{cases}
 h_N(\alpha)-1,& \gamma=\alpha\\
 h_N(\gamma), & \gamma\neq \alpha
 \end{cases}
\]
and
\[
  J^\alpha h_N (\gamma) = \begin{cases}
  h_N(\alpha)+1,& \gamma=\alpha\\
  h_N(\gamma),& \gamma\neq\alpha.
  \end{cases}
\]
Note that the transition $h_N\mapsto J^\alpha_\beta$ preserves the mass of the crystal, $m=\sum_{\alpha \in \mathbb{T}^d_N} h_N(\alpha)$.

Now that we have defined the transitions by which the crystal evolves we need to specify the rate at which those transitions occur.  To that end we first define the \emph{generalized coordination number}, $n(\alpha)$ for $\alpha\in \mathbb{T}^d_N$ by
\begin{multline}\label{gcn}
n_N(t,\alpha) =  \frac{1}{2}\sum_{i\leq d}V(\nabla^+_i J_\alpha h_N(t,\alpha)) - V(\nabla^+_i h_N(t,\alpha)) \\+ V(\nabla^-_i J_\alpha h_N(t,\alpha)) - V(\nabla^-_i h_N(t,\alpha)).
\end{multline}
One can think of $n(\alpha)$ as the (symmetrized) energy cost associated with removing a single atom from site $\alpha$ on the crystal surface.  

We will assume that the atom at site $\alpha$ breaks the bonds with its nearest neighbors at a rate that is exponential in the generalized coordination number.  Once those bonds are broken the atom chooses a neighboring site of $\alpha,$ for example $\beta$ with $|\beta -\alpha|=1$, uniformly and jumps there, i.e. $h_N\mapsto J^\beta_\alpha h_N.$
Since there are $2d$ sites $\beta$ with $|\beta-\alpha|=1,$ the rate of a transition
$
h_N\mapsto J_\alpha^\beta h_N
$
is
\[
r_N(t,\alpha) = \frac{1}{2d} e^{- 2 K n_N(t,\alpha)}.
\]
As with $h_N$ we will occasionally omit the $t$ argument in $n_N$ and $r_N.$

The above description of the evolution of the process $h_N$ is summarized by its generator $\mathcal{A}_N.$  Knowledge of the generator allows us to characterize the evolution of any function $f$ of the crystal surface by,
\[
f(h_N(t,\alpha))-f(h_N(0,\alpha)) = \int_0^t \left[\mathcal{A}_N f\right](s,\alpha)
+ M_f(t,\alpha)
\]
where $M_f(t,\alpha)$ is a random process with $M_f(0,\alpha)=0$ and whose expectation at time $t$ (over realizations of $h_N$) given the history of $h_N$ up to time $s\leq t$ is simply its value at time $s.$  In particular $\mathbf{E}\left[ M_f(t,\alpha)\right] = 0$ for all $t$ and $\alpha$ where $\mathbf{E}$ is used to denote the expectation over many realizations of the surface evolution from a particular initial profile.  For our process,
\begin{equation}\label{gen}
\mathcal{A}_N f (h_N) =  \sum_{\substack{\alpha,\beta\in\mathbb{T}^d_N \\ |\alpha-\beta|=1}} r_N(\alpha) \left( f(J_\alpha^\beta h_N) - f(h_N)\right).
\end{equation}
One can check that 
\[
\langle g\,(\mathcal{A}_N f)\rangle_N^m =  \sum_{h_N} g\,(\mathcal{A}_N f)\,p_N^m(h_N) = \sum_{h_N} f\,(\mathcal{A}_N g)\,p_N^m(h_N) = \langle f\,(\mathcal{A}_N g)\rangle_N^m,
\]
i.e. that $\mathcal{A}_N$ is self adjoint with respect to the $p^m_N$ weighted inner product.  The jump process defined by the rates above is reversible and ergodic with respect to $p_N^m.$

There are many possible choices for the rates (and corresponding definitions of the generalized coordination number) that would yield dynamics ergodic with respect to $p_N^m.$  What distinguishes our
particular choice (besides consistency with established models) is the fact that  the generalized coordination numbers defined in \eqref{gcn} are independent of the neighbor $\beta$ of $\alpha$ to which the surface atom at site
$\alpha$ will move.
  This structure is  motivated by our physical interpretation of the generalized coordination number as
the cost of breaking all bonds holding the surface atom at lattice site $\alpha.$  Once these bonds are all broken the atom is free to chose a neighbor of $\alpha$ uniformly.  This viewpoint is consistent with the classical description of chemical reaction rates in terms of energy barriers (see \cite{KDM}).
We could define an alternative generalized coordination number by replacing $J_\alpha$ in \eqref{gcn} by $J^\alpha.$  This new coordination number would also be independent of the   neighbor  to which the surface atom at site
$\alpha$ will move.  This generalized coordination number, however, would  measure the cost to attach the atom previously at site $\alpha$ at a neighboring site.   Such a choice does not appear to us to be physically motivated.

\begin{example}[SOS] Suppose $V( z) = |z|,$ which is the example considered in \cite{KDM}.  Then, 
\[
n_N(\alpha) + 2^{d-1} =  \sum_{\substack{\beta\in\mathbb{T}^d_N \\ |\alpha-\beta|=1}} \mathbf{1}_{(h_N(\alpha)\leq h_N(\beta))} 
\]
where 
\[
\mathbf{1}_{(h_N(\alpha)\leq h_N(\beta))} = \begin{cases} 1 & \text{if } h_N(\alpha)\leq h_N(\beta)\\
0 & otherwise
\end{cases}.
\]
In words,  up to an additive constant (which amounts to a time rescaling), the generalized coordination number is the number of neighbor bonds that need to be broken to free the atom at lattice site $\alpha.$
\end{example}
\begin{example}[discrete Gaussian model] Suppose $V(z) = z^2.$  Then
\[
n_N(\alpha) - 2d =  \sum_{i\leq d} \nabla^+_i h_N(\alpha) - \nabla^-_i h_N(\alpha),
\]
i.e. up to an additive constant, the generalized coordination number is the discrete Laplacian of the surface at lattice site $\alpha.$
\end{example}

In both of the examples above, one can view the generalized coordination number as a measure of the curvature of the surface near site $\alpha.$  The resulting rates treat positive and negative curvature very differently and one might expect, therefore, that surface regions of a positive curvature will evolve very differently from surface regions of similar but negative curvature.  One interesting conclusion that can be drawn from the results in the next section is that in the standard large crystal scaling limit this asymmetry vanishes while it is very apparent in the second scaling limit that we consider.

\section{PDE limits}
\label{sec:PDElimits}

Before we can specify the PDE limits that we consider, we need to define the relevant scaling limits.
The first scaling regime is standard.  For reasons that will be explained later we refer to this regime as the {\it smooth} scaling limit.
For any function $f: [0,\infty)\times \mathbb{T}^d_N \rightarrow \mathbb{R}$ we define the projections
$\bar f_N: [0,\infty)\times [0,1]^d \rightarrow \mathbb{R}$ by 
\begin{equation}\label{smoothscaling}
\bar f_N(t,x) = N^{-1}  f(N^4 t, \alpha)\qquad \text{for}\qquad  N x \in \bigcap_{i=1}^d \left[ \alpha_i - \frac{1}{2}, \alpha_i+\frac{1}{2}\right).  
\end{equation}
In Sections \ref{s:numerics} and \ref{s:smoothkrug} we argue that
$\bar h_N(t,x)$ converges to the solution of the PDE
\begin{equation}\label{smoothpde}
\partial_t h= - K \Delta \text{div}\left[\sigma_D(\nabla h)\right]
\end{equation}
where, for $u\in \mathbb{R}^d,$ the surface tension $\sigma_D(u)$ is the derivative of the surface free-energy,
\begin{equation}\label{free-energy1}
\mathcal{F}_D(u) = \frac{1}{K}\sup_{\sigma\in \mathbb{R}^d }\left\{ \sigma^\text{\tiny T} u  - \log \Psi_D (\sigma)\right\} 
\end{equation}
with
\[
\Psi_D(\sigma) = 
 {\sum_{z\in \mathbb{Z}^d} e^{-K \sum_{i\leq d}V(z_i) + \sigma^\text{\tiny T} z }}.
 \] 
 Notice that 
 \begin{equation}
 \label{eqn:sigmaD}
 \sigma_D(u)= \nabla \mathcal{F}_D (u)
 \end{equation} 
 is the value of $\sigma$ at which the optimum in 
 \eqref{free-energy1} is attained.  The surface tension satisfies
 \[
u = \left[\nabla \Psi_D\right](\sigma(u)) =  \frac{\sum_{z\in \mathbb{Z}^d} z\, e^{-K \sum_{i\leq d}V(z_i) +   K\sigma_D^\text{\tiny T} z}}
 {\sum_{z\in \mathbb{Z}^d} e^{-K\sum_{i\leq d} V(z_i) + K\sigma_D^\text{\tiny T} z}}
 \]
 i.e. $\sigma_D(u)$ is exactly the value of the external field $\sigma$ that shifts the mean of the distribution 
 \[
 \frac{ e^{-K\sum_{i\leq d}V(z_i) + K\sigma^\text{\tiny T} z }}{\Psi_D(\sigma)}
 \]
 to $u.$  
 
 In one spatial dimension, with $V(z) = |z|,$ the PDE \eqref{smoothpde} with the $\sigma_D$ just defined is exactly the PDE suggested in \cite{KDM}.
 However, it differs in the definition of the surface tension from the PDE identified in \cite{vvedensky}.  As mentioned above,  the discrepancy with \cite{vvedensky} is due to an additional, small lattice constant approximation made in that work.  In that approximation the height variable becomes continuous, $h_N(t,\alpha)\in \mathbb{R}$ and is governed by the over-damped Langevin equation 
 \begin{multline}\label{langevin}
 d h_N(t,\alpha) = -\sum_{\substack{\beta\in \mathbb{T}^d_N\\ i\leq d}} L_{\alpha \beta} \left(V'(\nabla^+_i h_N(t,\beta))-V'(\nabla^-_i h_N(t,\beta))\right)\, dt\\ + \sqrt{2K} \sum_{\beta\in \mathbb{T}^d_N} \left(\sqrt{-L}\right)_{\alpha \beta}\, dW(t,\beta)
 \end{multline}
 where $L$ is the discrete Laplacian matrix on the lattice,
 \[
 L_{\alpha \beta}  = \begin{cases} 1 & \text{if } |\alpha-\beta | = 1\\
  -2 d & \text{if } \alpha = \beta\\
  0 & \text{otherwise}
  \end{cases},
  \]
  $\sqrt{-L}$ is the square root of the positive semi-definite matrix $-L,$
 and $W$ is an independent Brownian motion for each $\alpha.$
 
 The continuum limit of the diffusion in \eqref{langevin} was studied rigorously by Nishikawa in \cite{Nishikawa} where it is shown that
 $\bar h_N(t,x)$ converges to the solution of the PDE
\begin{equation}\label{smoothpdeC}
\partial_t h= -K \Delta \text{div}\left[\sigma_C(\nabla h)\right]
\end{equation}
with surface tension 
\begin{equation}
\label{eqn:sigmaC}
\sigma_C(u) = \nabla \mathcal{F}_C(u),
\end{equation} where
 \begin{equation}\label{free-energy2}
\mathcal{F}_C(u) = \frac{1}{K} \sup_{\sigma\in \mathbb{R}^d}\left\{\sigma^\text{\tiny T} u - \log \Psi_C (\sigma)\right\} 
\end{equation}
and
\[
\Psi_C(\sigma) = 
 {\int e^{-K \sum_{i\leq d}V(u_i) + K\sigma^\text{\tiny T} u} du }.
 \] 
 
 Clearly the surface tensions $\sigma_D$ and $\sigma_C$ are different and so, therefore, are the solutions of the corresponding PDE \eqref{smoothpde} and \eqref{smoothpdeC}.  We explore this difference numerically in the next section.  That discussion has two primary outcomes.  On the one hand, we are able to conclusively discern that the PDE with $\sigma_D$ is a better representation of the crystal surface evolution in this scaling regime.  On the other hand, that distinction is very difficult to diagnose as the solutions of the PDE with $\sigma_D$ and $\sigma_C$ are extremely close.  For more discussion of this issue see the next section.
 
Before moving on to a description of our second scaling limit, we point out that  one very interesting qualitative feature of the PDE evolution in \eqref{smoothpde} is that if the potential $V$ is symmetric and the initial condition is symmetric (respectively skew-symmetric) about $x=0,$ then the solution of the
PDE \eqref{smoothpde} is symmetric (respectively skew-symmetric) at all times.  This is in sharp contrast to the behavior of the 
KMC model itself where positive curvature and negative curvature have very different effects on the rates.  It is however, consistent with the over-damped Langevin microscopic model \eqref{langevin}.

 Our second scaling regime is less standard. We refer to it as the {\it rough} scaling limit.
We will assume that for some $p>1$ the potential $V$ is homogenous of degree $p,$ i.e.
\begin{equation}\label{Vscaling}
V(z) =  \kappa^{-p}V(\kappa z)
\end{equation} 
for all $\kappa>0.$
As before let $\sigma_D(u) = \nabla \mathcal{F}_D(u)$ where
 \begin{equation*}
\mathcal{F}_D(u) = \frac{1}{K}\sup_{\sigma\in \mathbb{R}^d}\left\{ \sigma^\text{\tiny T} u - \log \Psi_D (\sigma)\right\} .
\end{equation*}  
Our second PDE limit will require that we characterize the behavior of $\sigma_D(u)$ for very large $u.$  More precisely we need to consider the limit  $\kappa^{1-p}\sigma_D(\kappa u)$ as  $\kappa$ grows very large.   As we will argue in Section \ref{s:roughkrug}, we expect that the limit of $\kappa^{1-p}\sigma_D(\kappa u)$
exists and that
\begin{equation}\label{barsigma}
\bar \sigma(u) = \lim_{\kappa \rightarrow \infty} \kappa^{1-p}\sigma_D(\kappa u) = \nabla V(u).
\end{equation}

Now set
\[
q = \frac{p}{p-1}
\]
and, for any function $f: [0,\infty)\times \mathbb{T}^d_N \rightarrow \mathbb{R},$ define the projections
$\bar f_N: [0,\infty)\times [0,1]^d \rightarrow \mathbb{R}$ by 
\begin{equation}\label{roughscaling}
\bar f_N(t,x) = N^{-q}  f(N^{q+2} t, \alpha)\qquad \text{for}\qquad  N x \in \bigcap_{i=1}^d \left[ \alpha_i - \frac{1}{2}, \alpha_i+\frac{1}{2}\right).  
\end{equation}
In Sections \ref{s:numerics} and \ref{s:roughkrug} we argue that
$\bar h_N(t,x)$ converges to the solution of the PDE
\begin{equation}\label{roughpde}
\partial_t h = \Delta  \exp\left(-\text{div}\left[\bar\sigma(\nabla h)\right]\right).
\end{equation}
This PDE is very similar to the one identified in the last pages of \cite{KDM} in $1+1$ dimensions with $V(z)=|z|,$ differing only in the definition of the surface tension.   

In some respects this non-standard scaling limit is the more interesting regime.  It retains many of the interesting features of the microscopic system that are lost in the more standard scaling regime defined by \eqref{smoothscaling}.  
For example, we have remarked above that if $V$ is symmetric about 0 and the initial surface is symmetric (or antisymmetric) about $x=0,$ then the solution to \eqref{smoothpde} is symmetric (or antisymmetric) about $x=0$ for all time.  This does not hold for equation $\eqref{roughpde}$ and certainly does not hold for the microscopic evolution.  On the other hand, a PDE very similar to \eqref{smoothpde} can be derived from \eqref{roughpde} by considering profiles with very small curvature.  This  explains our use of the terms smooth and rough to differentiate our scaling limits.  Indeed the rough regime can be thought of as describing very large, rapidly varying surfaces.
 In the next section we will numerically explore the features of the two scaling limits more carefully.

\section{Numerical Experiments and Discussion}
\label{s:numerics}

We now provide a numerical comparison of our microscale and macroscale models.  We will place particular emphasis on diagnosing the correct form of the surface tensions appearing in \eqref{smoothpde} and in \eqref{roughpde}.   In the smooth scaling limit giving rise to \eqref{smoothpde} this means differentiating between $\sigma_D$ and $\sigma_C$ defined in \eqref{eqn:sigmaD} and \eqref{eqn:sigmaC} above.  As we will show, straightforward comparisons of the corresponding numerical solutions of the PDE does not clearly reveal the correct choice.   In the second scaling limit giving rise to \eqref{roughpde} we will numerically explore the effect of the limit in \eqref{barsigma} defining $\bar \sigma.$  The above comparisons will be performed in $1+1$ dimensions.  We will conclude this section by showing the results of several simulations in $2+1$ (2 spatial dimensions and 1 time dimension) dimensions that demonstrate that the qualitative behavior of the systems does not seem to be effected by the dimension.   Unless otherwise noted, the initial profile for both the PDE simulation and the rescaled microscopic evolution is $\sin(2\pi x)$ in 1+1 dimensions and $\sin(2\pi x)\sin(2\pi y)$ in 2+1 dimensions.
Results will only be shown for $K=1.5$ as we did not find that the value of $K$ had any effect (in the $1+1$ or $2+1$ dimensional cases) on the qualitative features that we remark on below.

\begin{figure}
\centering
\includegraphics[width=2.5in]{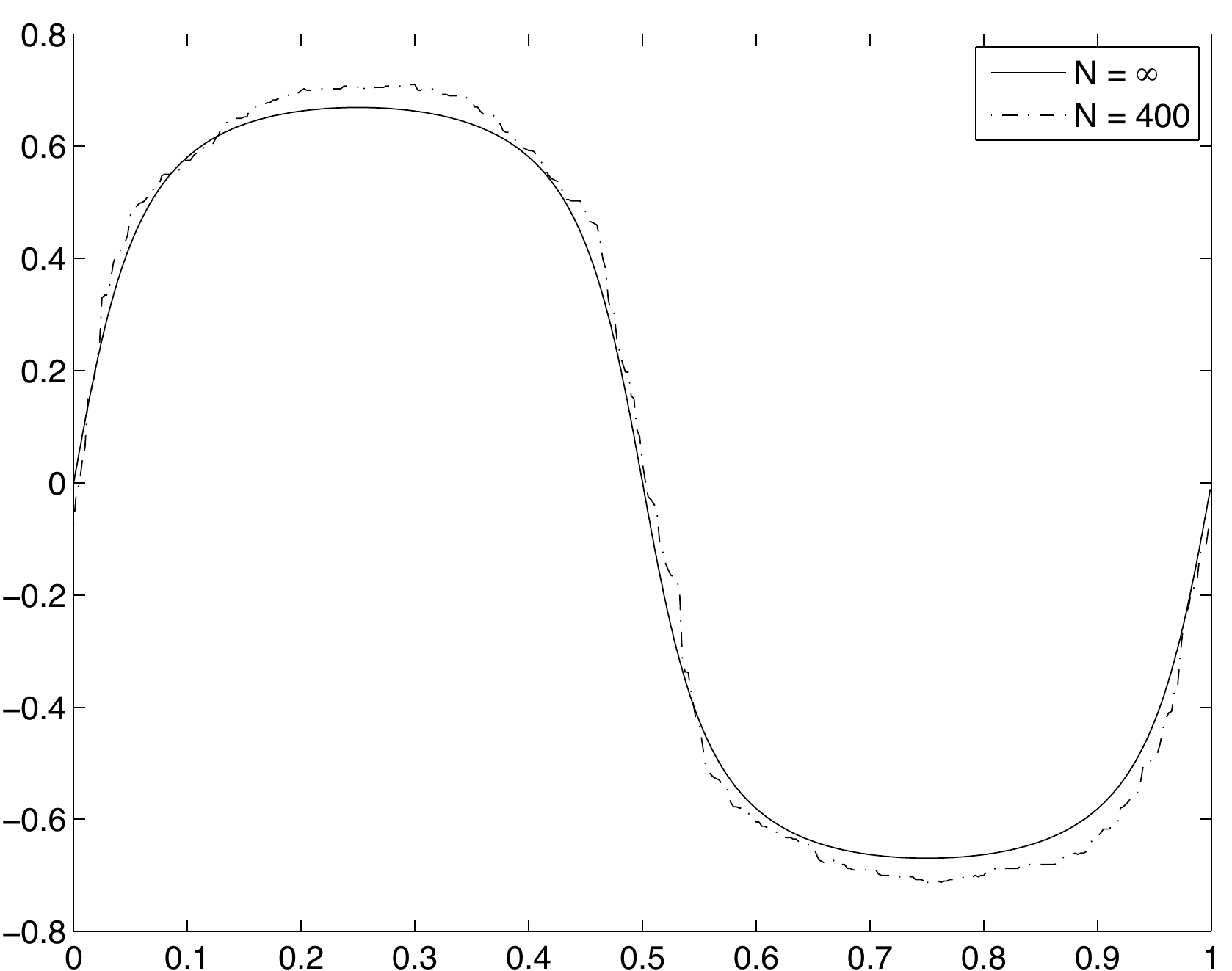}
\includegraphics[width=2.5in]{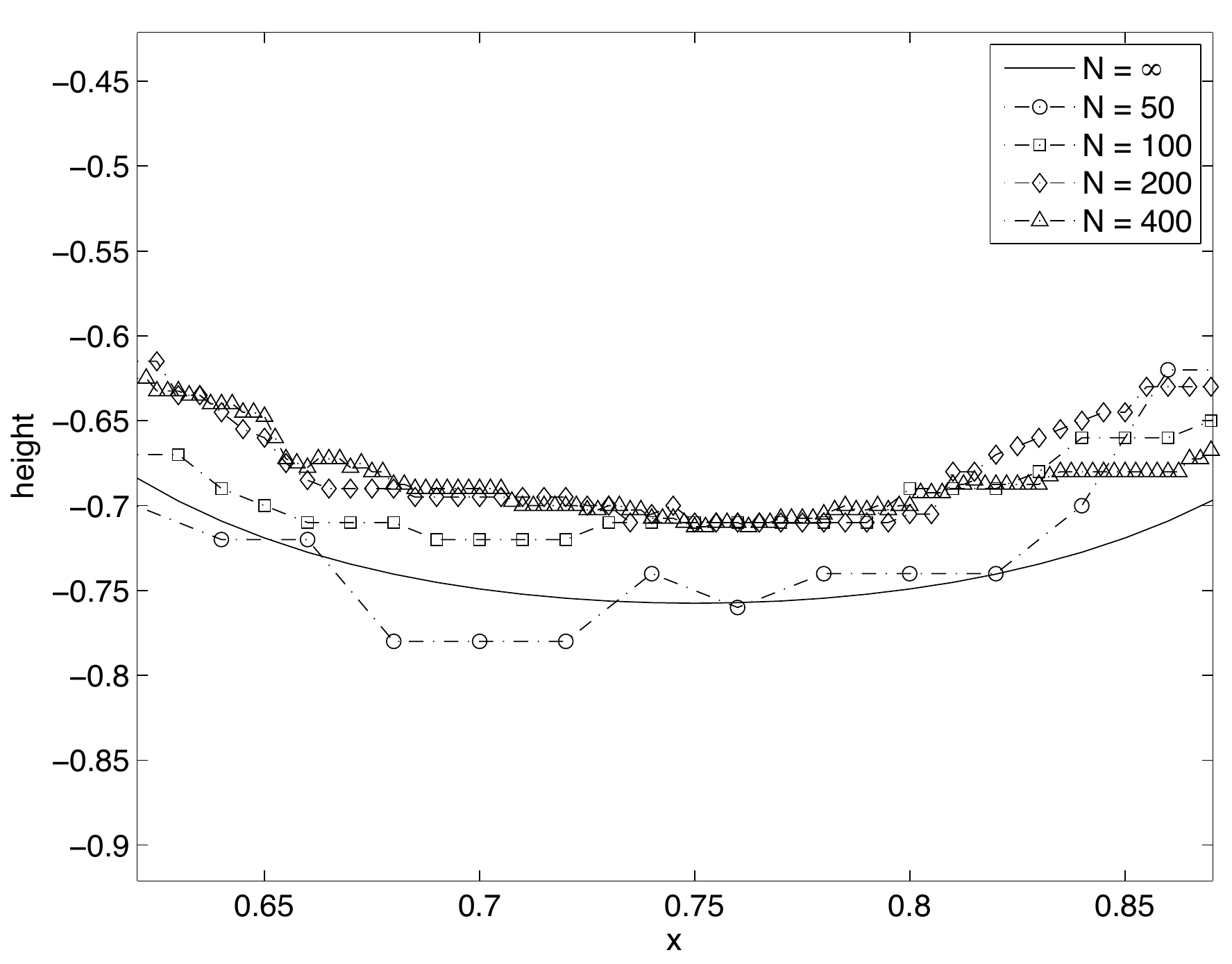}
\caption{Comparison of the solution of  PDE  \eqref{smoothpde} (labeled $N=\infty$) with $V(z)=|z|$ for $T = 10^{-3}$ at $K=1.5$ to the appropriately rescaled microscopic profile with $N = 400$  and a blow-up near the minimum for $N=50,100,200,400$ in 1+1 dimensions.}
\label{fig:p1smooth}
\end{figure}

We begin by demonstrating the convergence, in the smooth scaling limit (defined in \eqref{smoothscaling}) of microscopic model to the solution of the PDE \eqref{smoothpde}.  Figure \ref{fig:p1smooth} compares the rescaled microscopic evolution ($\bar h_N$ defined as in \eqref{smoothscaling}) at time $T=10^{-3}$ to the solution of \eqref{smoothpde} at the same time for various values of $N.$  Here $V(x) = |x|,$ i.e. Figure \ref{fig:p1smooth} represents the SOS model.    Since $\sigma_D(u)$ is the inverse of $K^{-1}\nabla \log \Psi_D(\sigma)$ (which can be easily approximated numerically), we can compute and store the value of $\sigma_D$ at a set of points and interpolate as needed.   The PDE simulations are all run at a fine enough resolution to be considered fully converged for the purposes of these comparisons.  The agreement between the rescaled microscopic profile for $N=400$ and the solution to the PDE is on the order of $0.1$.  Since the rescaled microscopic profile has noise features on roughly the same scale we attribute the remaining mismatch to the effects of a finite $N.$  Unfortunately simulations of the microscopic system at large enough $N$ to realize convergence are not feasible.  Below we will describe an alternative experiment that allows us to compare the microscopic evolution with larger $N$ to the PDE.  For other choices of $V$ the picture is much more clear.

\begin{figure}
\centering
\includegraphics[width=2.5in]{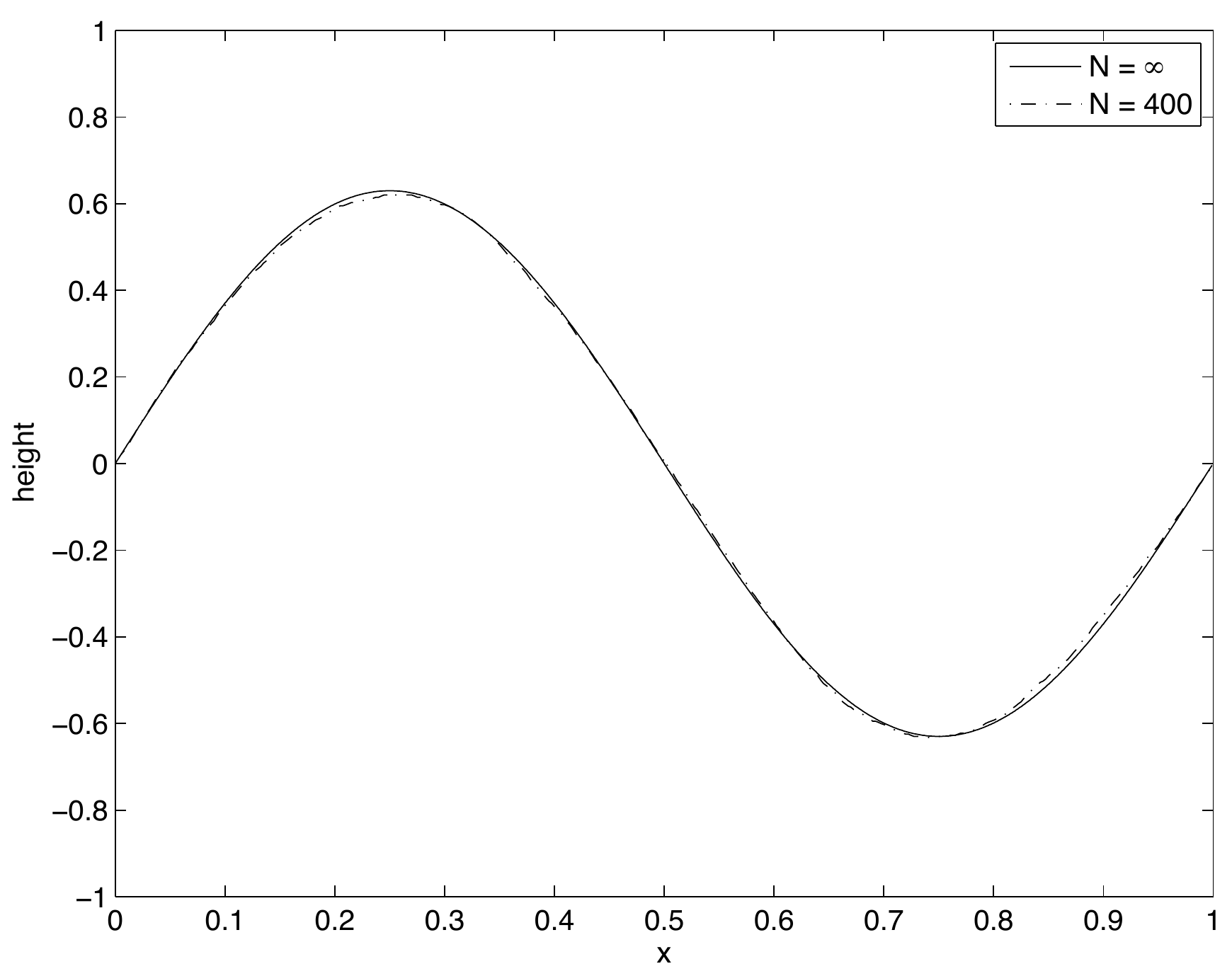}
\includegraphics[width=2.5in]{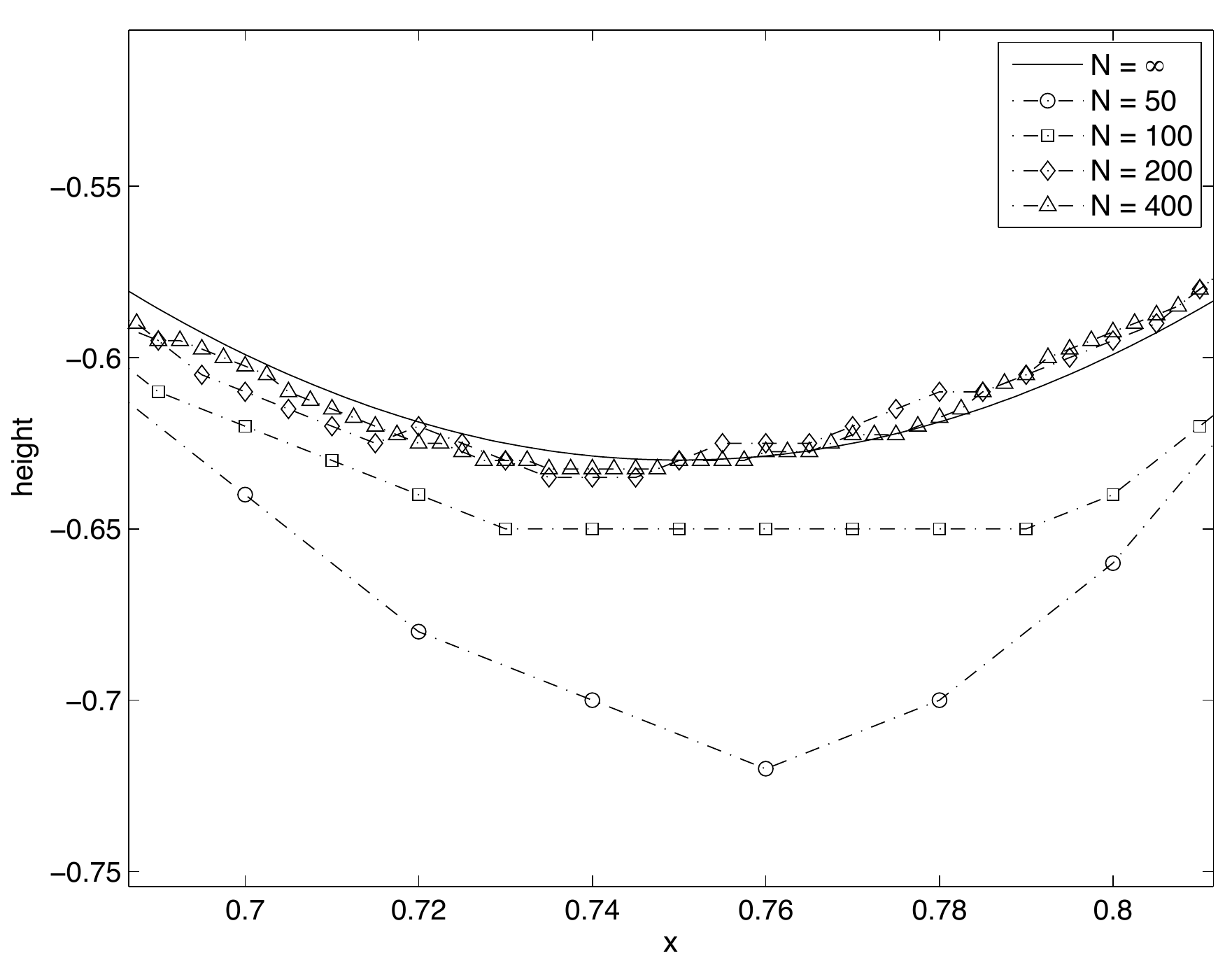}
\caption{Comparison of the solution of PDE \eqref{smoothpde}  (labeled $N=\infty$) with $V(z)=z^2$ for $T = 2\times 10^{-4}$ at $K=1.5$ to the appropriately rescaled  microscopic profile with $N = 200$  and a blow-up near the minimum for $N=50,100,200$ in 1+1 dimensions.}
\label{fig:p2smooth}
\end{figure}

Figure \ref{fig:p2smooth} compares the rescaled microscopic evolution with $V(z) = z^2$ to the solution of \eqref{smoothpde} with the same $V.$  Both profiles are plotted at $T= 2\times 10^{-4}.$  Here the agreement between the PDE solution and the rescaled microscopic profile is more convincing.

\begin{figure}
\centering
\includegraphics[width=2.5in]{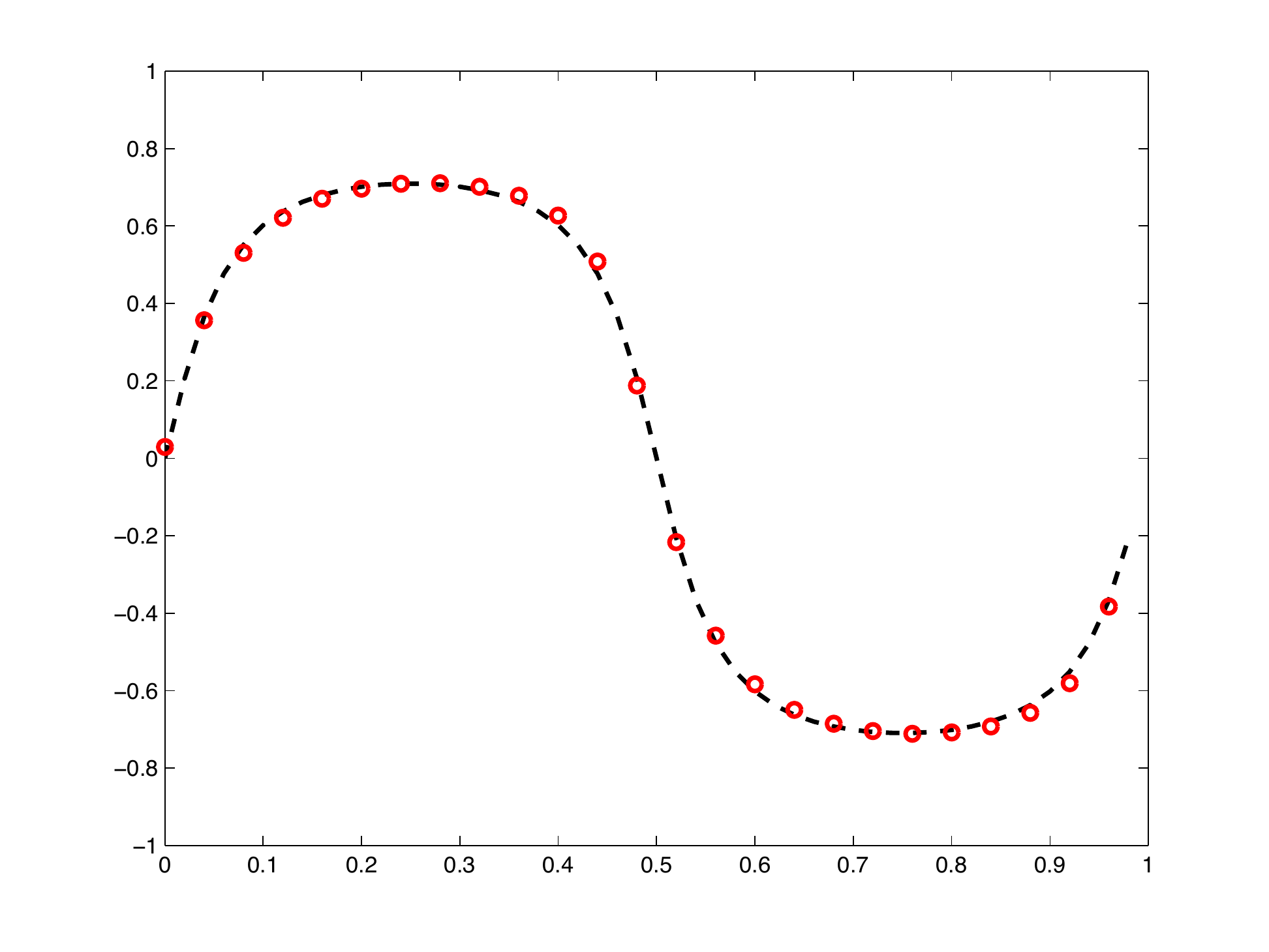}
\includegraphics[width=2.5in]{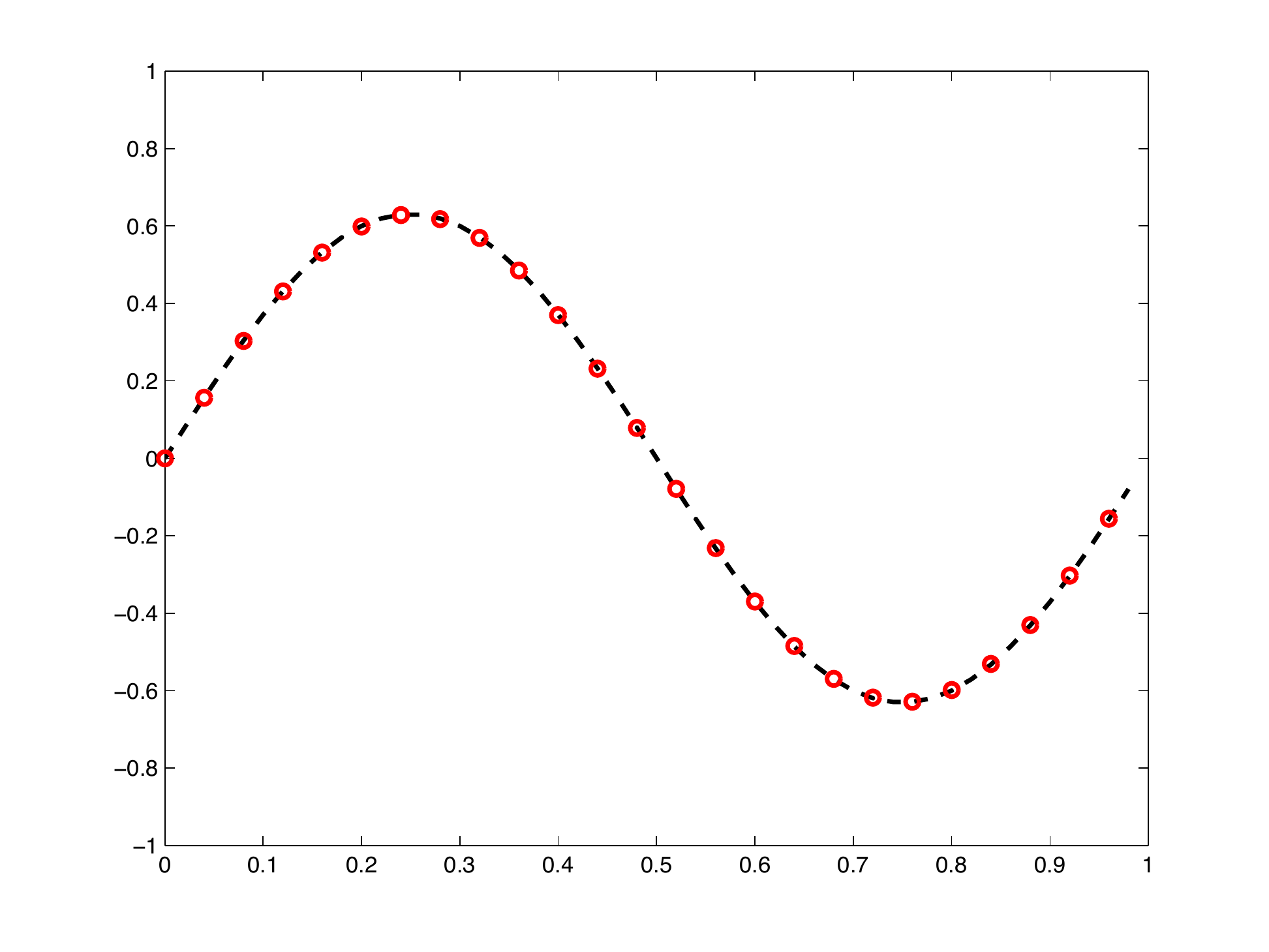}
\caption{Comparison of the solution of PDE \eqref{smoothpde} with the surface tensions  $\sigma_C$ and $\sigma_D$ for $V(z)=|z|$, $T=10^{-3}$ (left) and $V(z)=z^2$, $T = 2\times 10^{-4}$ (right)  in $1+1$ dimensions.}
\label{fig:sigmacomp}
\end{figure}

We have remarked above that numerically differentiating between different definitions of the surface tension ($\sigma_D$ or $\sigma_C$)  is difficult.  Figure \ref{fig:sigmacomp} demonstrates this fact.  For the potentials $V(z)=|z|$ and $V(z)= z^2$ it shows that the solutions of the \eqref{smoothpde} with the two different definitions of the surface tension are very similar.  Unfortunately, the difference between the two solutions is far below the resolution that we are able achieve with our microscopic simulations in reasonable time and we are not able to resolve the ambiguity 
by straightforward simulation with a large $N.$  We therefore appeal to the generator $\mathcal{A}_N$ defined in \eqref{gen}.  The generator satisfies
\begin{align*}
&\frac{\mathbf{E}\left[ h_N(T,x)\right] -  h_N(0,x)}{T} =  \frac{1}{T}\mathbf{E}\left[ \int_0^T\mathcal{A}_N h_N\, (s,x)\,ds \right]\\
 &\hspace{2cm}= \frac{1}{T} \mathbf{E}\Bigg[\int_0^T \sum_{\substack{\beta\in \mathbb{T}^d_N\\ |\beta-\alpha|=1}} r_N( s, \beta) - r_N( s, \alpha)\,ds \Bigg]
\end{align*}
where 
\[
N x \in \bigcap_{i=1}^d \left[ \alpha_i - \frac{1}{2}, \alpha_i+\frac{1}{2}\right).
\]
In terms of $\bar h_N$ this can be rewritten as
\begin{equation*}
\frac{\mathbf{E}\left[ \bar h_N(T,x)\right] -  \bar h_N(0,x)}{T} =  
\frac{N^3 }{T} \mathbf{E}\Bigg[\int_0^T \sum_{\substack{\beta\in \mathbb{T}^d_N\\ |\beta-\alpha|=1}} r_N(N^4 s, \beta) - r_N(N^4 s, \alpha)\,ds \Bigg] .
\end{equation*}

\begin{figure}
\centering
\includegraphics[width=3in]{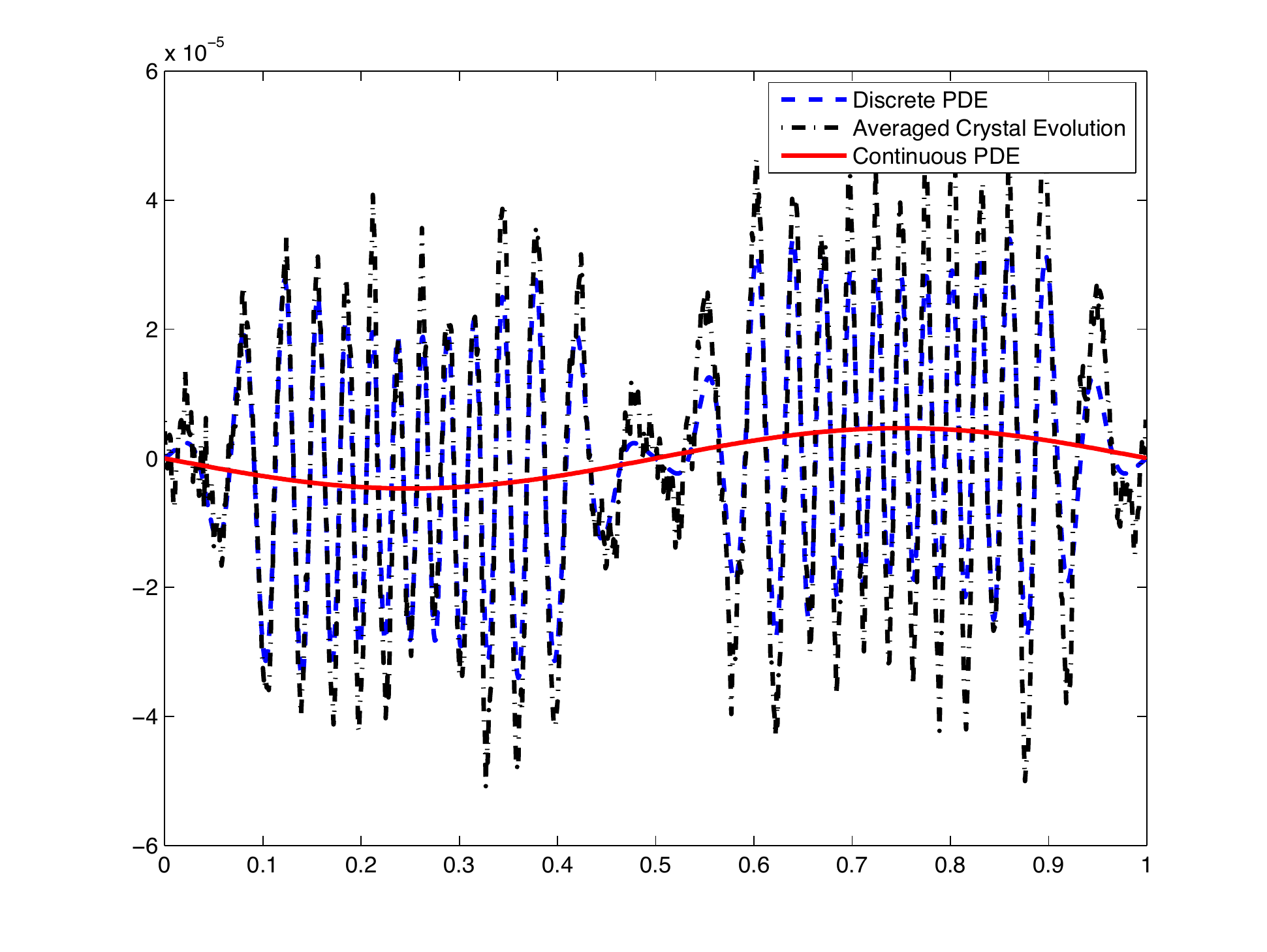}
\caption{Comparison of  the left and right hand sides of \eqref{ifsigmaD} and \eqref{ifsigmaC} for $V(z)=z^2$ at $T=2\times 10^9$ with $N=1000.$  In the legend, ``Discrete PDE'' refers to equation \eqref{smoothpde} and ``Continuous PDE'' refers to equation \eqref{smoothpdeC}.}
\label{fig:p2smoothcompsigma}
\end{figure}

If we choose a value of $T$ in the range $N^{-4} \ll T \ll 1$ (i.e.  a $T$ that is large on the length scale of the microscopic evolution but short on the time scale of the PDE evolution), then we should find that
\begin{align*}
& \partial_t \mathbf{E}\left[\bar h_N(T,x)\right] \approx \\
& \hspace{2cm} \frac{N^3}{T} \mathbf{E}\Bigg[\int_0^T \sum_{\substack{\beta\in \mathbb{T}^d_N\\ |\beta-\alpha|=1}} r_N(N^4 s, \beta) - r_N(N^4 s, \alpha)\,ds\Bigg].
\end{align*}
 Thus if $\bar h_N$ is approaching a deterministic function solving \eqref{smoothpde}, then  we should have
\begin{align}
& -K \Delta \text{div}\left[ \sigma_D(\nabla \mathbf{E}\left[\bar h_N(T,\cdot)\right])\right]  \approx \notag \\
& \hspace{2cm}  \frac{N^3}{T} \mathbf{E}\Bigg[\int_0^T \sum_{\substack{\beta\in \mathbb{T}^d_N\\ |\beta-\alpha|=1}} r_N(N^4 s , \beta) - r_N( N^4 s, \alpha)\,ds\Bigg] . \label{ifsigmaD}
\end{align}
If the limit of $\bar h_N$ solves \eqref{smoothpdeC} instead then we should have that 
\begin{align}
& -K\Delta \text{div}\left[ \sigma_C(\nabla \mathbf{E}\left[\bar h_N(T,\cdot)\right])\right] \approx \notag \\
& \hspace{2cm}  \frac{N^3}{T} \mathbf{E}\Bigg[\int_0^T \sum_{\substack{\beta\in \mathbb{T}^d_N\\ |\beta-\alpha|=1}} r_N(N^4 s, \beta) - r_N(N^4 s, \alpha)\,ds\Bigg] .\label{ifsigmaC}
\end{align}
The random variable inside the expectation on the right hand side of the last display has very large variance (especially when $N$ is large and $T$ is small) and computing the expectation requires a very large number of independent simulations of the microscopic model.  Fortunately, and unlike direct simulation of the system for long times, the simulation of many independent short trajectories of the system is a trivially parallelizable task.  Using the {\it Killdevil} cluster at UNC we were able to run $2\times 10^7$ sample trajectories with $N=1000$ and $T = 2\times 10^{-9}$ (corresponding to a microscopic evolution time of $2\times 1000^4\times 10^{-9}= 2\times 10^3$) and average the resulting realizations  of 
\[
\frac{N^3}{T} \int_0^T \sum_{\substack{\beta\in \mathbb{T}^d_N\\ |\beta-\alpha|=1}} r_N(N^4 s, \beta) - r_N(N^4 s, \alpha) \,ds.
\]
Note that the time integral above can be computed exactly.
The sample average is compared to the right hand side of the PDEs \eqref{smoothpde} and \eqref{smoothpdeC} in Figure \ref{fig:p2smoothcompsigma}.  The agreement with \eqref{smoothpde} is clearly superior to the agreement with \eqref{smoothpdeC}, indicating that the correct definition of the surface tension is $\sigma_D.$

\begin{figure}
\centering
\includegraphics[width=2.5in]{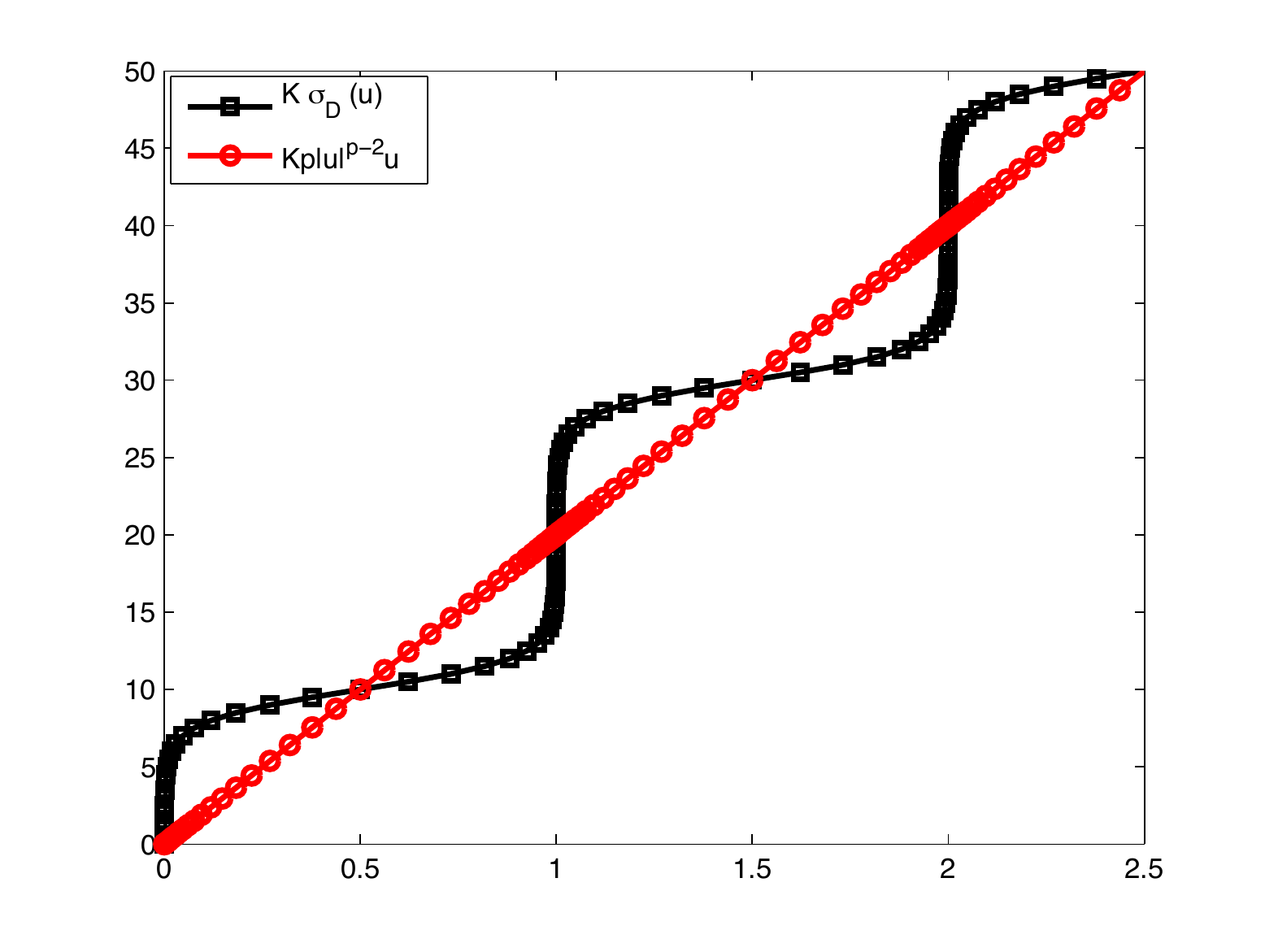}
\includegraphics[width=2.5in]{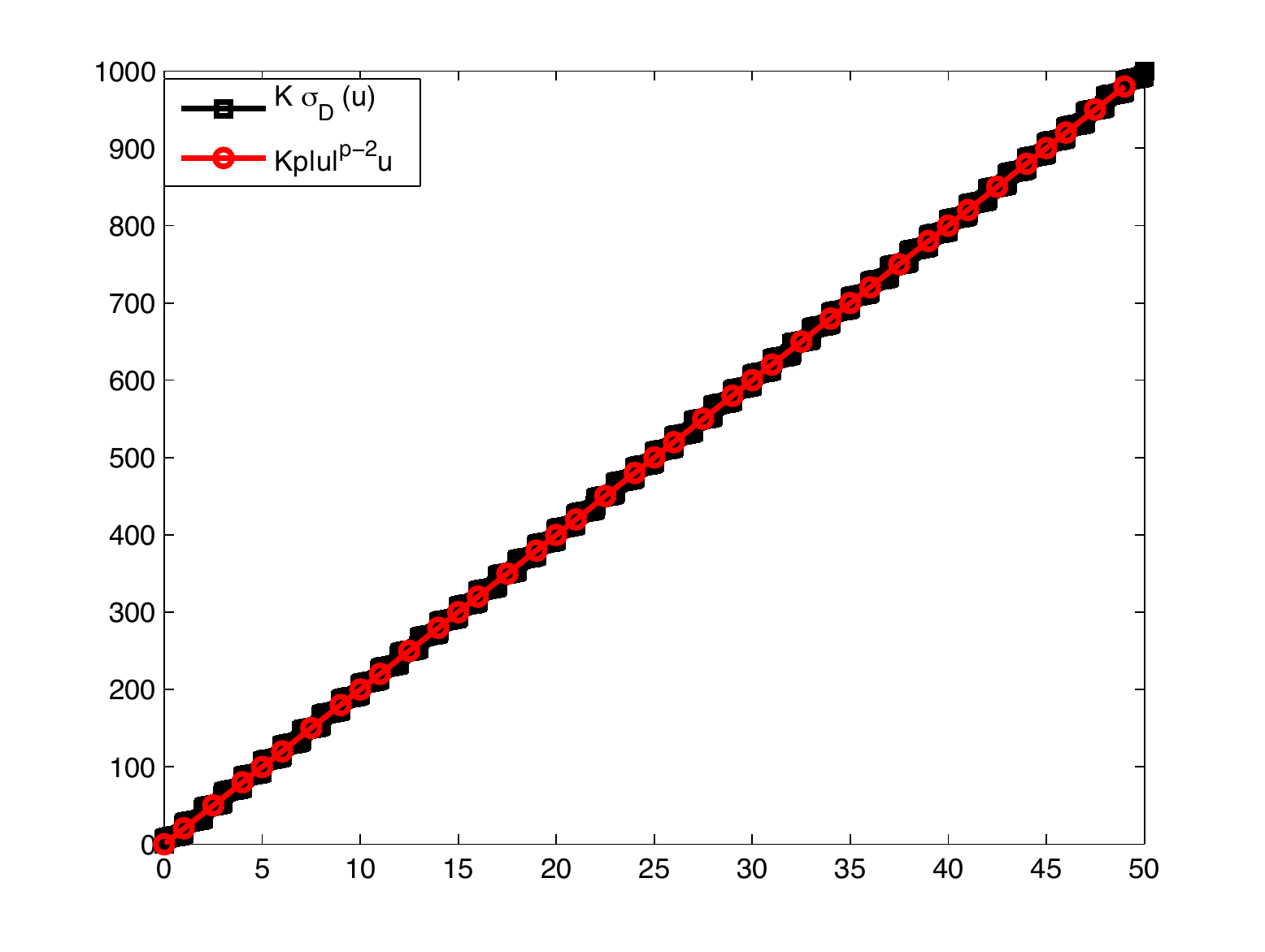}
\caption{Comparison of $K\bar \sigma (u)$ to $K\sigma_D(u)$ for $V(z)=z^2$ ($p=2$ in the legend) and $K =10$ at different scales to demonstrate the large scale behavior of $\bar \sigma.$}
\label{fig:p2smoothK10}
\end{figure}

\begin{figure}
\centering
\includegraphics[width=2.5in]{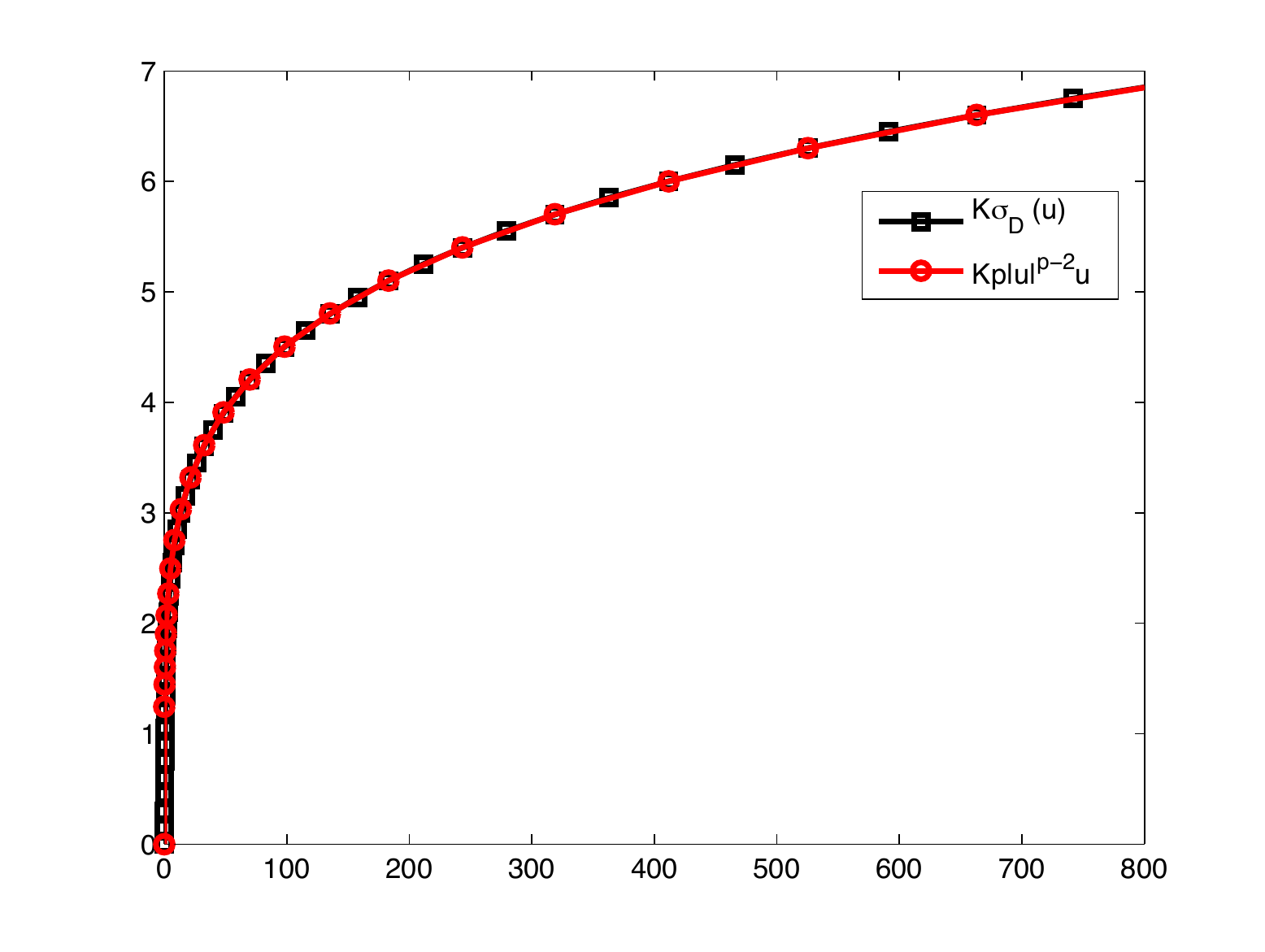}
\caption{Comparison of $K \sigma_D (u)$ to $K\bar \sigma(u) = Kp |u|^{p-2} u$ for $V(z)=|z|^p$ with $p = 1.2$ and $K =1.5$  to demonstrate the scaling law.}
\label{fig:sigmaKplots}
\end{figure}

Before moving to the convergence in the rough scaling limit (defined in \eqref{roughscaling}) of the microscopic model to the solution of the PDE \eqref{roughpde}, let us consider the definition $\bar \sigma $ in \eqref{barsigma}.  In Figures \ref{fig:p2smoothK10}
and \ref{fig:sigmaKplots} we plot $K\bar \sigma$ against $K\sigma_D$ for $V(z) = |z|^p$ with several values of $K$ and $p>1$ .  Notice that for these potentials, $\bar\sigma(u) = \lim_{\kappa\rightarrow \infty} \kappa^{1-p}\sigma_D(\kappa u)$ is effectively a smoothed version of $\sigma_D.$

\begin{figure}
\centering
\includegraphics[width=2.5in]{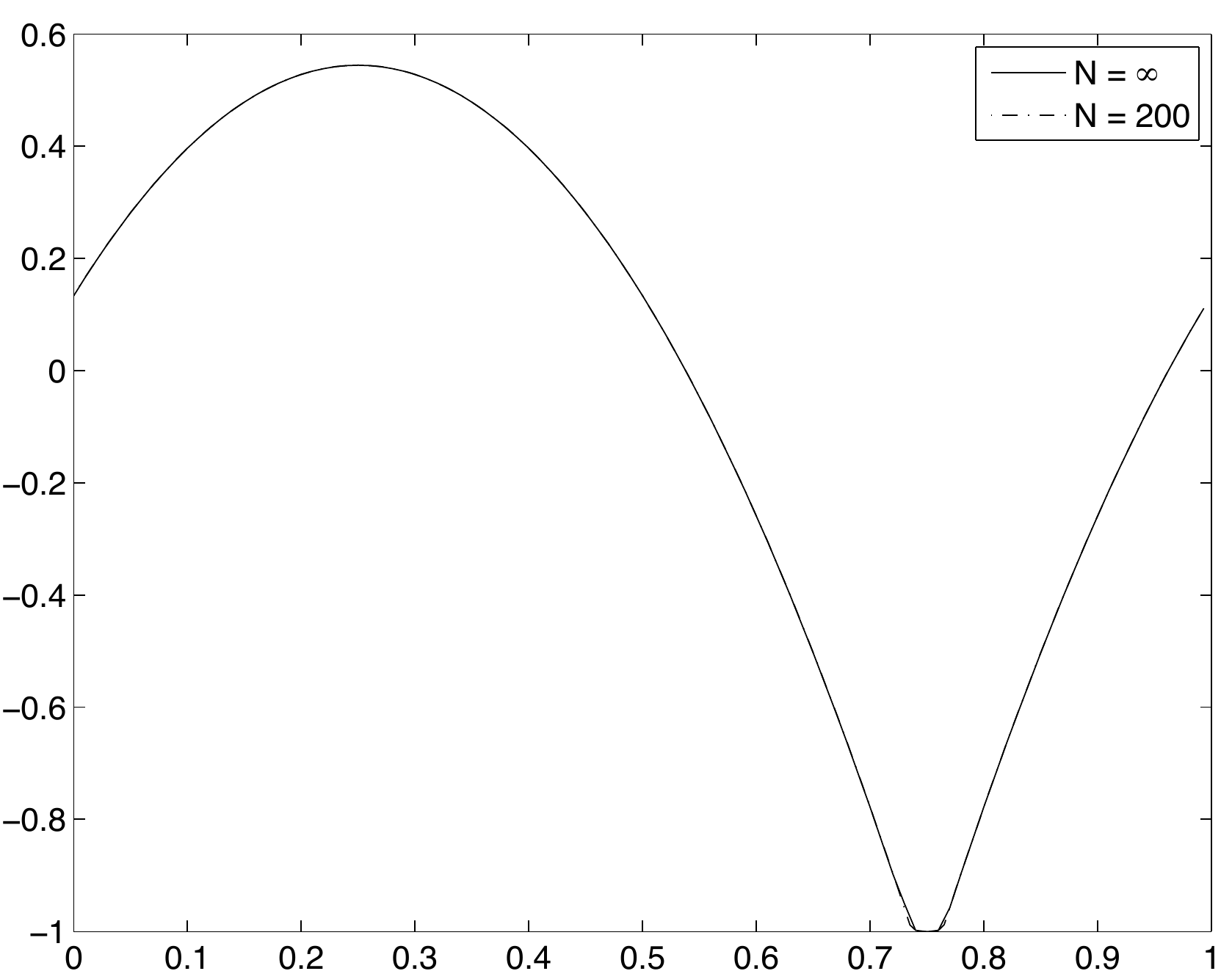}
\includegraphics[width=2.5in]{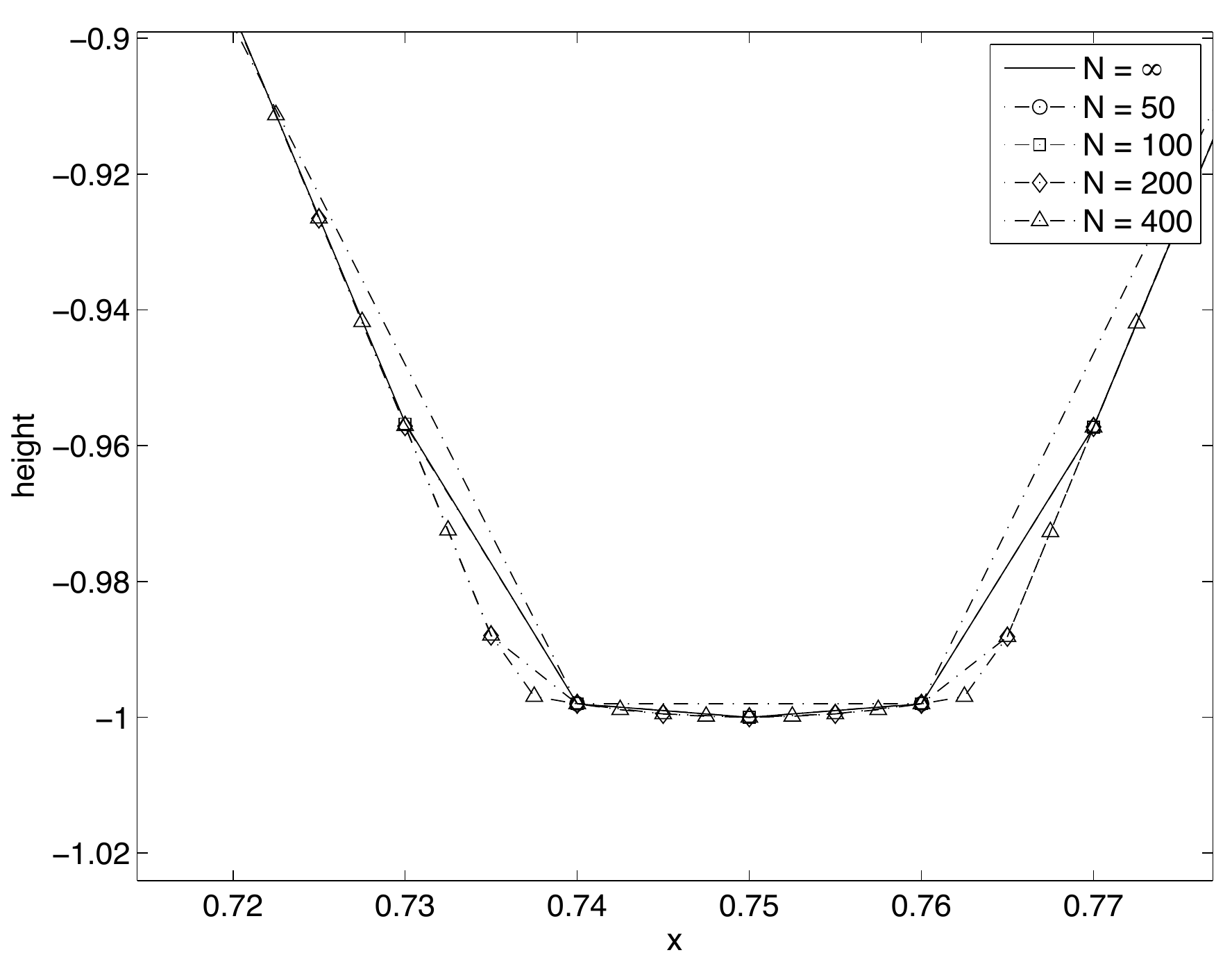}
\caption{Comparison of the  PDE \eqref{roughpde} (labeled $N=\infty$)  solution for $V(z)=z^2$ and $K = 1.5$ at $T = 10^{-20}$ to the appropriately rescaled microscopic profile with $N = 200$  and a blow-up near the minimum for $N=50,100,200,400$.}
\label{fig:p2rough}
\end{figure}

Now let us discuss the convergence of the microscopic system in the rough scaling limit.  Below we will present results only for $V(z) = z^2.$  We tested other potentials of the form $|z|^p$ for $p>1$ and found the qualitative behavior to be exactly the same as for $p=2.$ Figure \ref{fig:p2rough} compares the rescaled microscopic evolution ($\bar h_N$ defined as in \eqref{roughscaling}) at time $T=10^{-25}$ to the solution of \eqref{roughpde} at the same time for $N=50.$ Clearly the two surfaces agree well.  Note that the symmetry between the behavior of the peak and the valley that was present in the smooth scaling limit are not present here.  This scaling limit retains the microscopic model's asymmetry in the behavior of convex and concave regions of the surface.  

\begin{figure}
\centering
\includegraphics[width=2.5in]{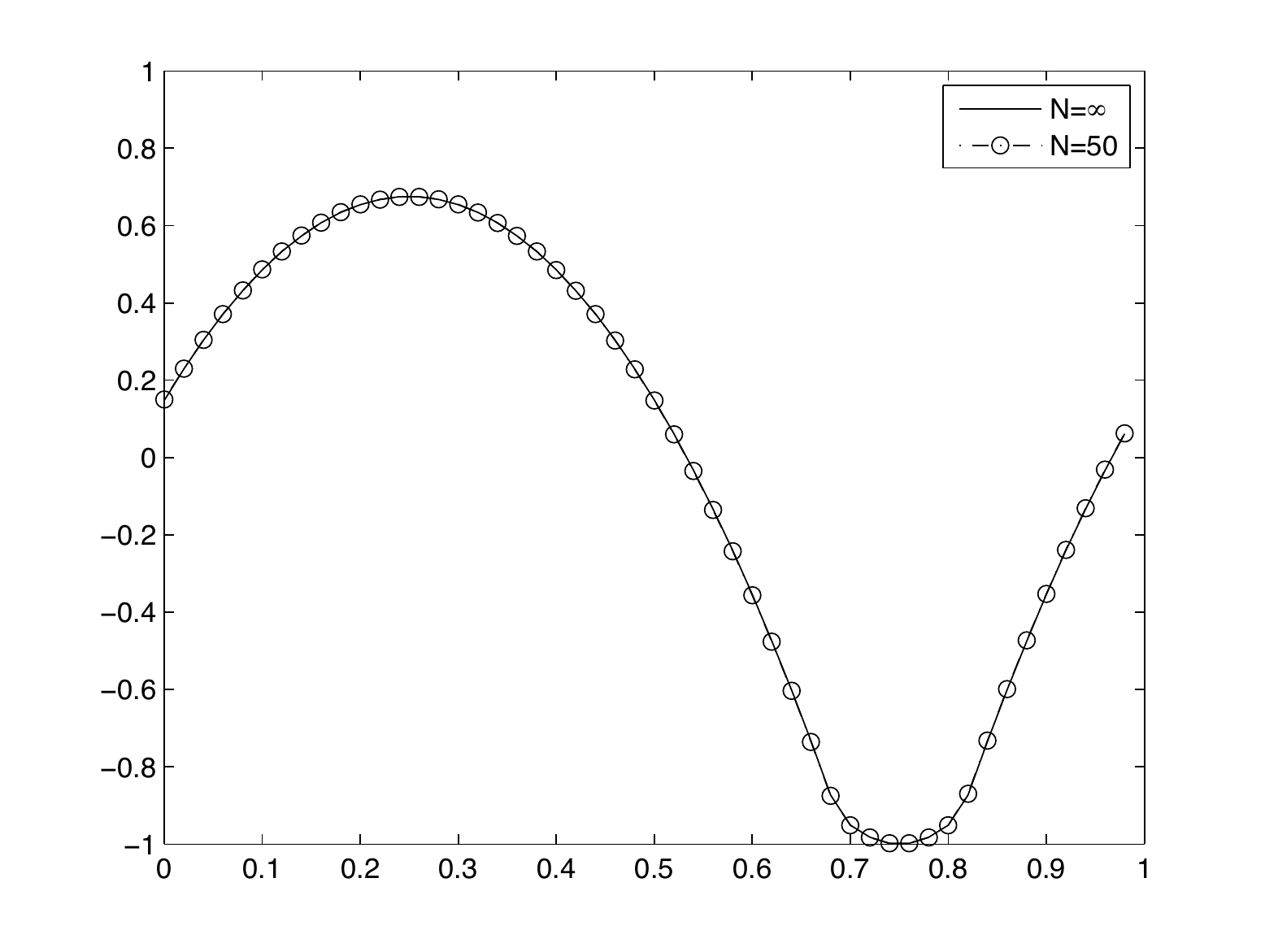}
\caption{Comparison of the solution of PDE \eqref{roughpde} (labeled $N=\infty$) to the appropriately rescaled  microscopic profile with $N=50$ for $K=1.5$ and $V(z)=z^2$ at $T = 10^{-25}$.}
\label{fig:p2rough1}
\end{figure}

Integrating both systems a bit further we observe another interesting feature of this rough scaling limit that does not appear to be present in the smooth scaling limit.  Figure \ref{fig:p2rough1} compares the rescaled microscopic evolution ($\bar h_N$ defined as in \eqref{roughscaling}) at time $T=10^{-20}$ to the solution of \eqref{roughpde} at the same time for various values of $N.$  Again, agreement between the rescaled microscopic model and the PDE solution is clear.  Now the surfaces have formed a non-smooth spike in the valley centered at $x=0.75.$  In the rough scaling limit the crystal appears to form singularities in regions of convexity, unlike the relatively smooth profiles generated in the smooth scaling limit.  In these simulations we chose $V(z) = z^2$ which corresponds to $\bar \sigma(u) = V'(u) =2 u.$   We investigated other potentials of the form $V(z) = |z|^p$ for $p>1$ and found the qualitative behavior to be generic ($p=1$ is not allowed in this scaling limit).

\begin{figure}
\centering
\includegraphics[width=3in]{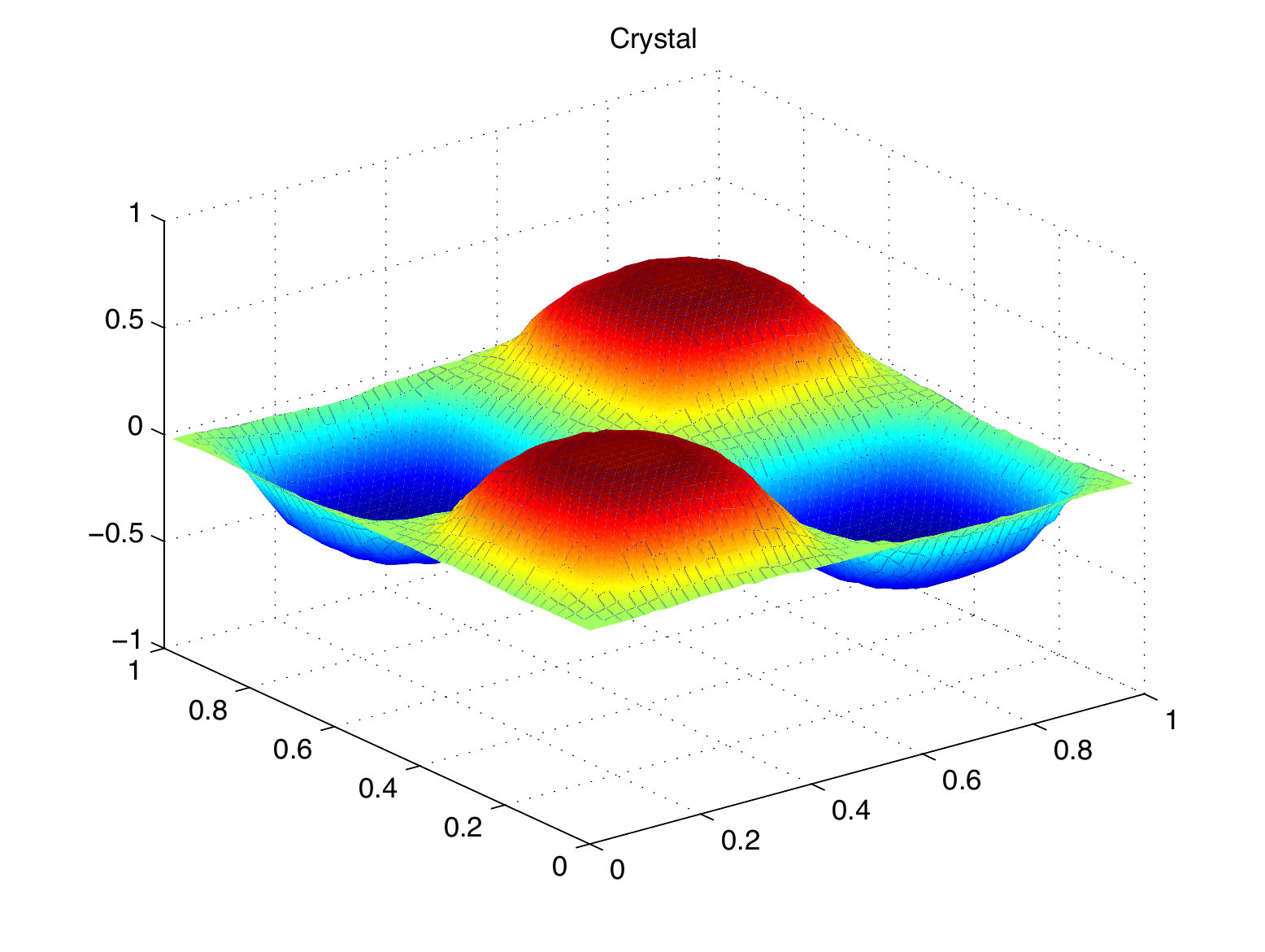} 
\includegraphics[width=3in]{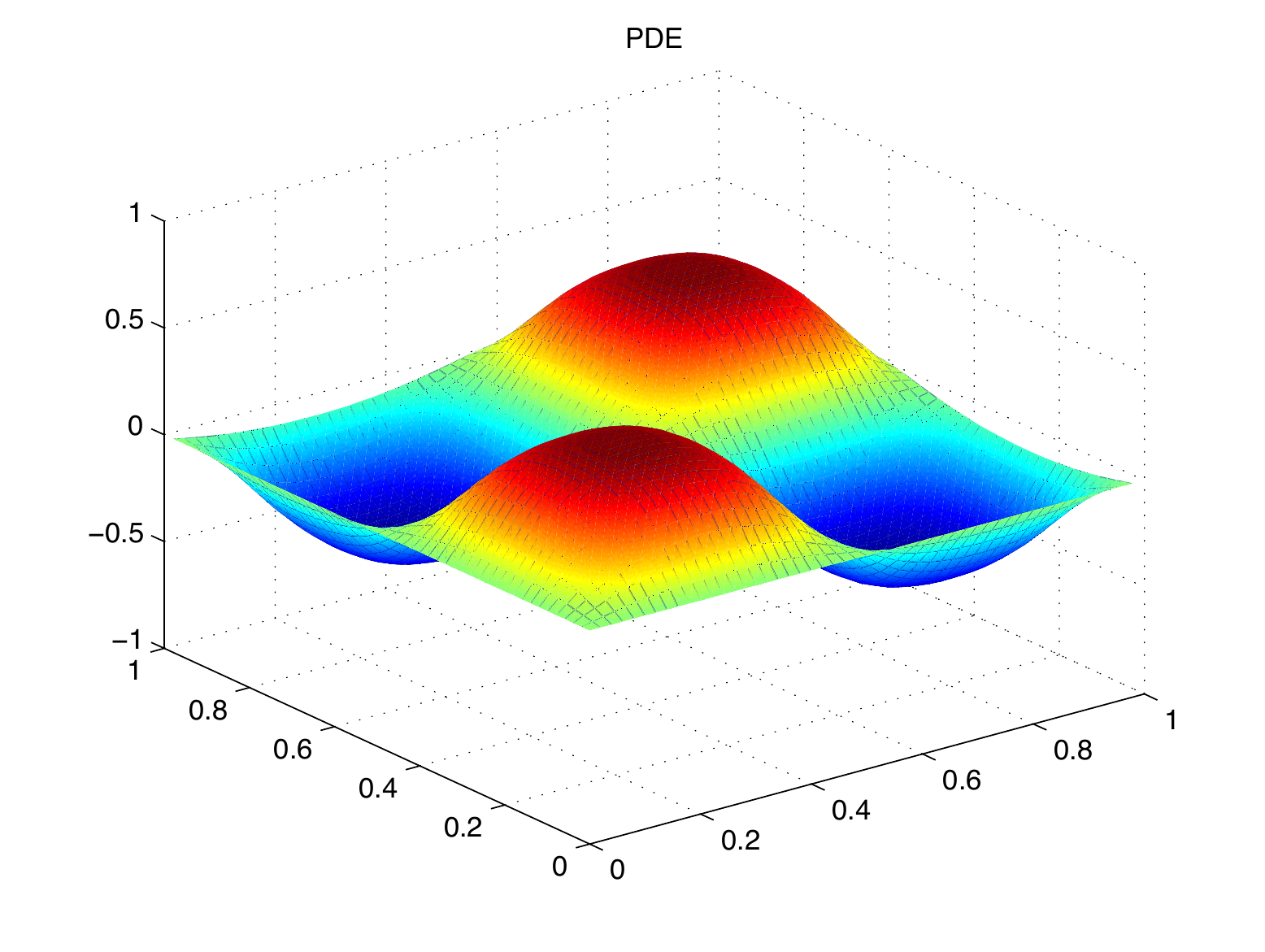} \\
\includegraphics[width=3in]{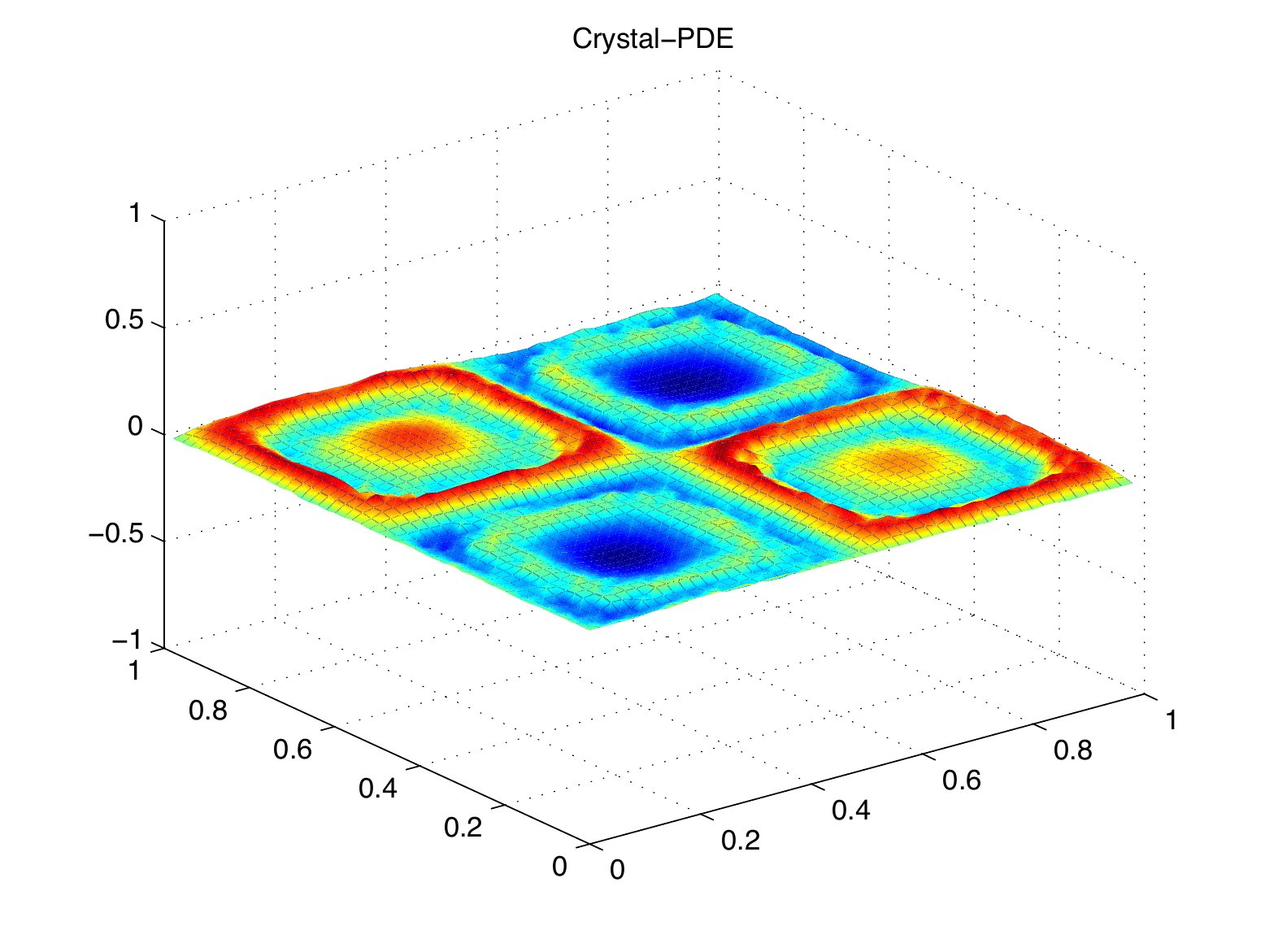}
\caption{The microscopic profile in the smooth scaling (top), the solution of PDE \eqref{smoothpde}  (middle), and the difference between the two (bottom) in 2+1 dimensions for $K=1.5$ with $V(z)=|z|$ at $T=10^{-3}$.  The maximum of the difference between the rescaled microscopic profile and the PDE solution is roughly $10^{-1}.$ }
\label{fig:p1smooth2d}
\end{figure}

\begin{figure}
\centering
\includegraphics[width=3in]{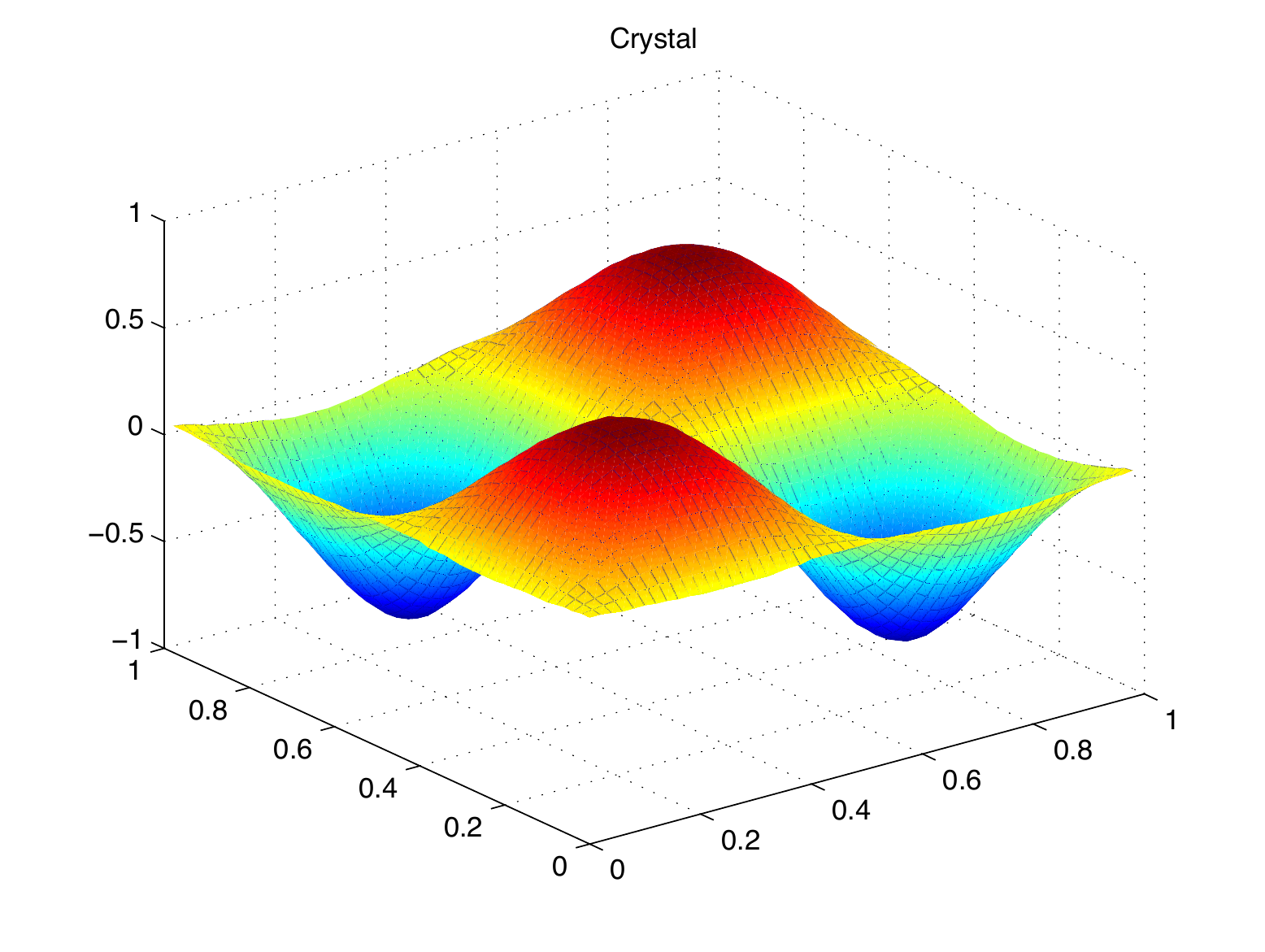} 
\includegraphics[width=3in]{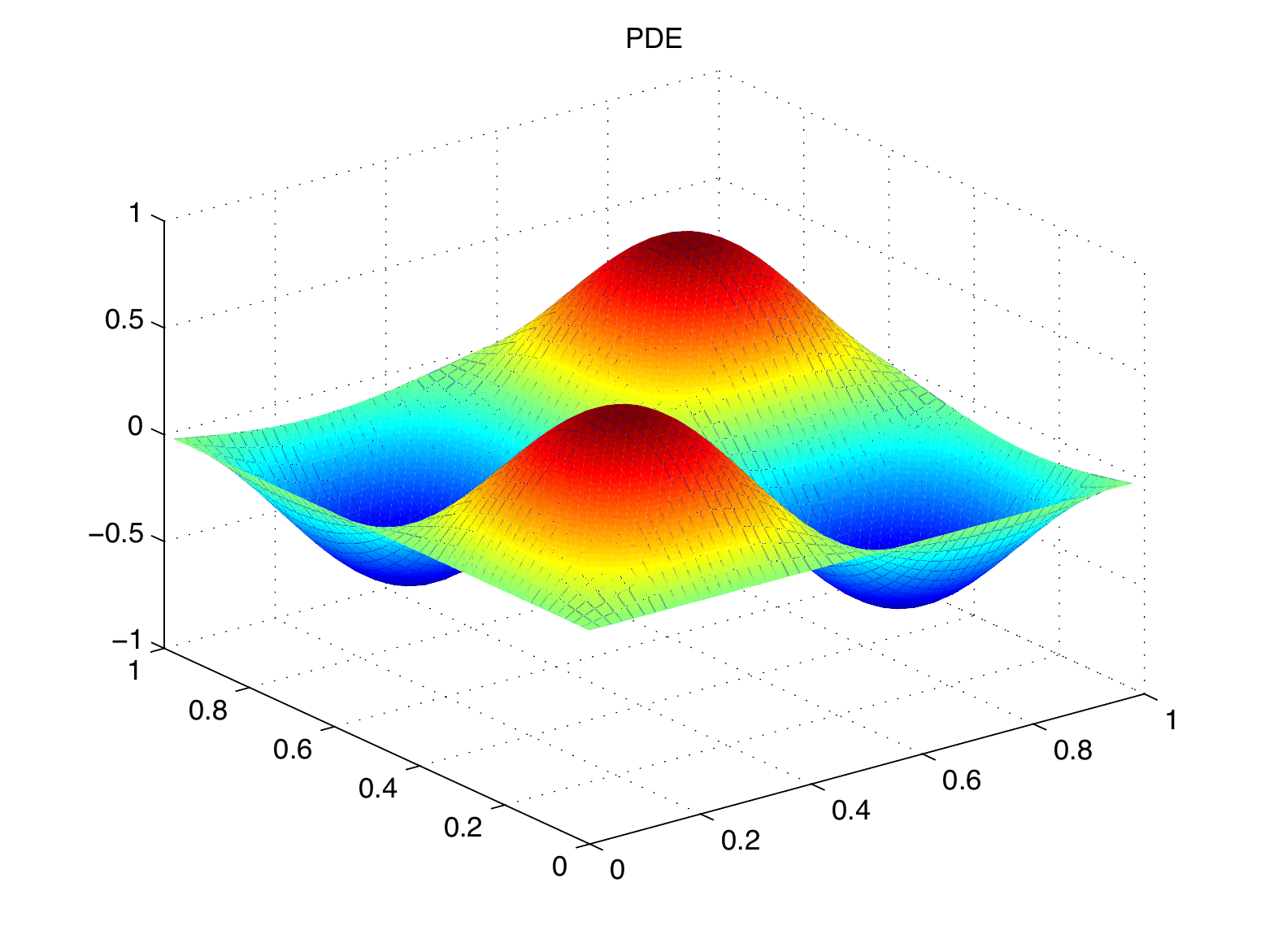} \\
\includegraphics[width=3in]{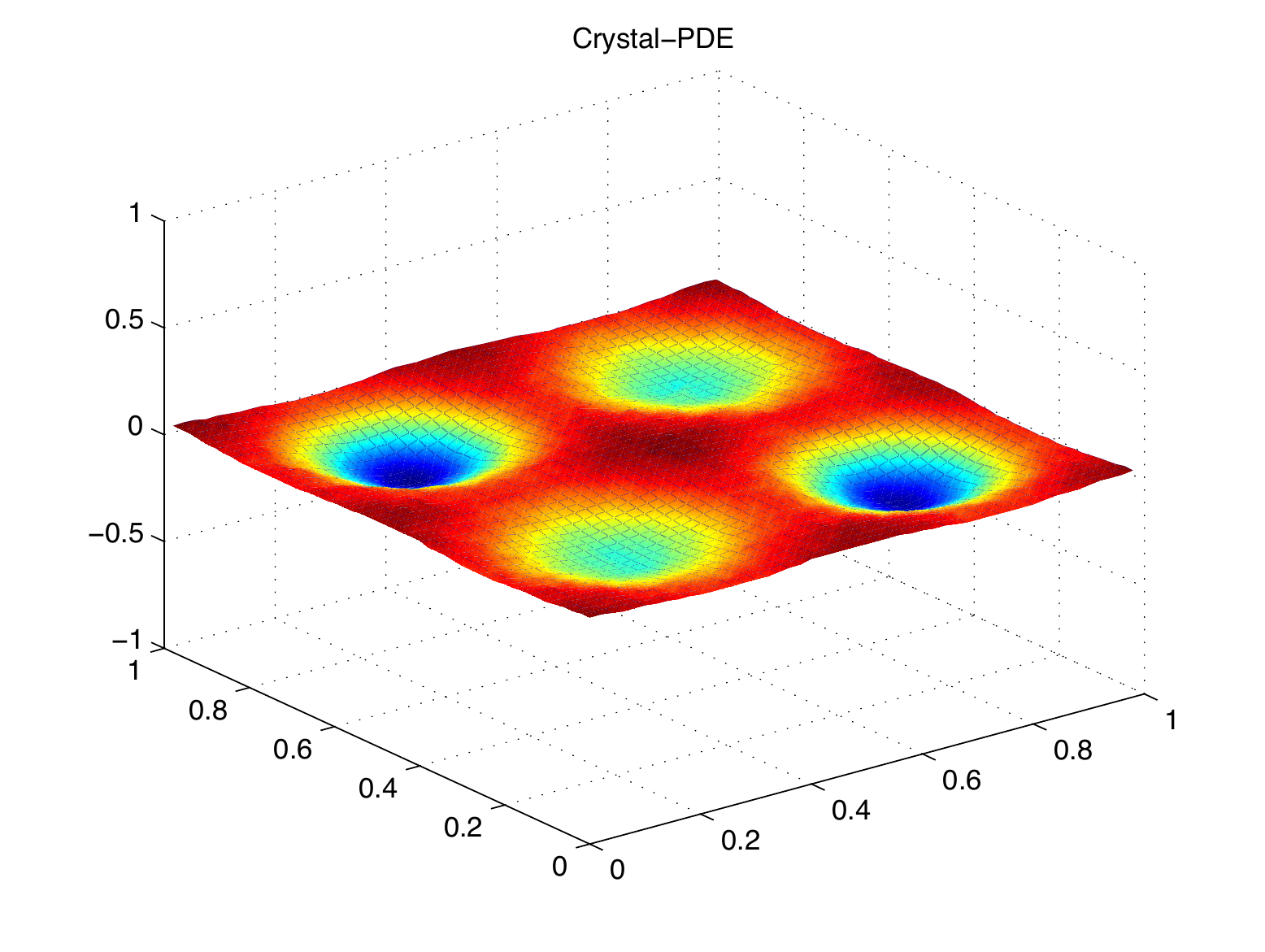}
\caption{The microscopic profile in the smooth scaling (top), the solution of PDE \eqref{smoothpde} (middle), and the difference between the two (bottom) in 2+1  for $K=1.5$ with $V(z)=z^2$ at  $T=10^{-4}$.  The maximum of the difference between the rescaled microscopic profile and the PDE solution is roughly $10^{-2}.$}
\label{fig:p2smooth2d}
\end{figure}

\begin{figure}
\centering
\includegraphics[width=3.0in]{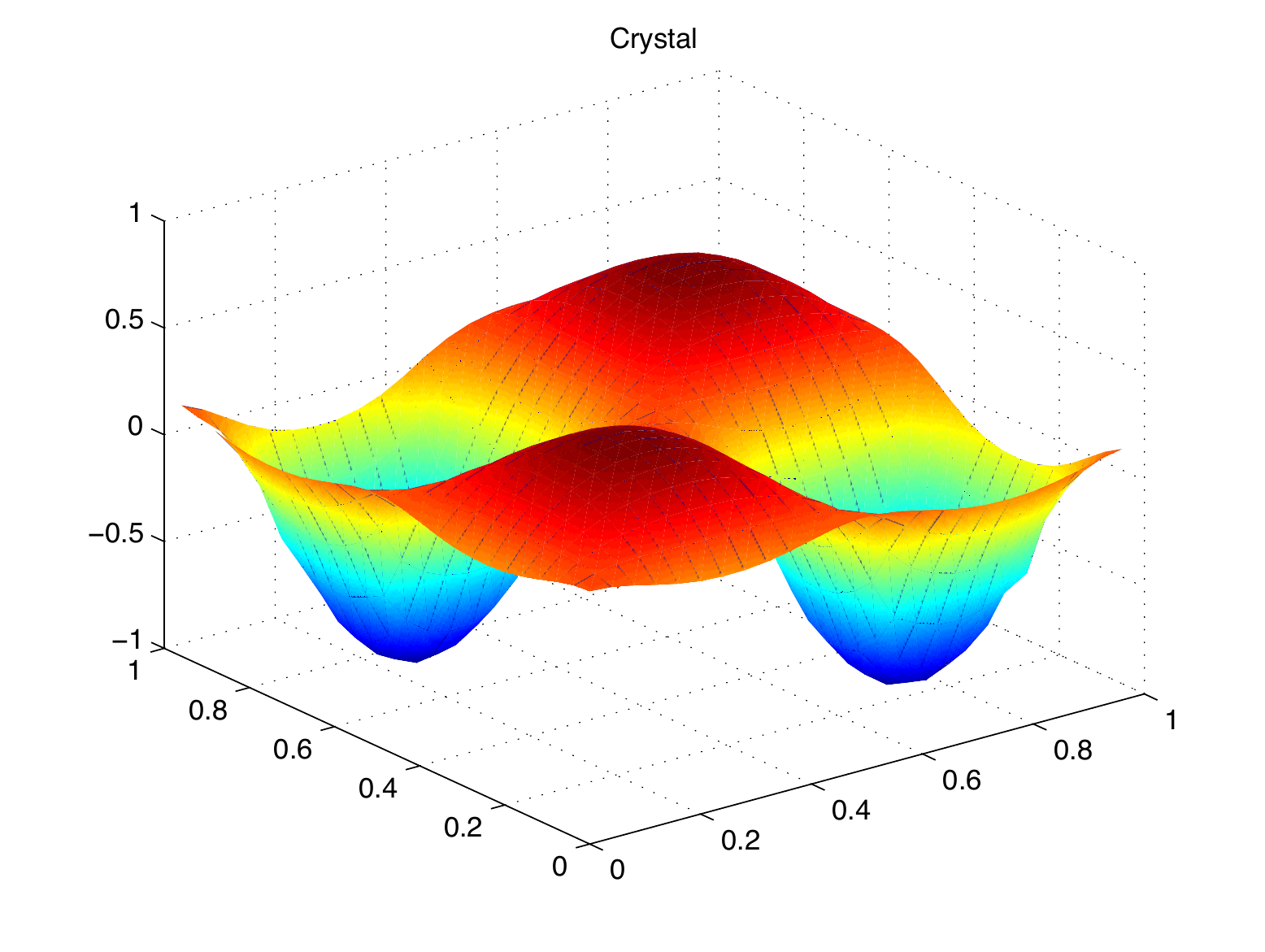} 
\includegraphics[width=3.0in]{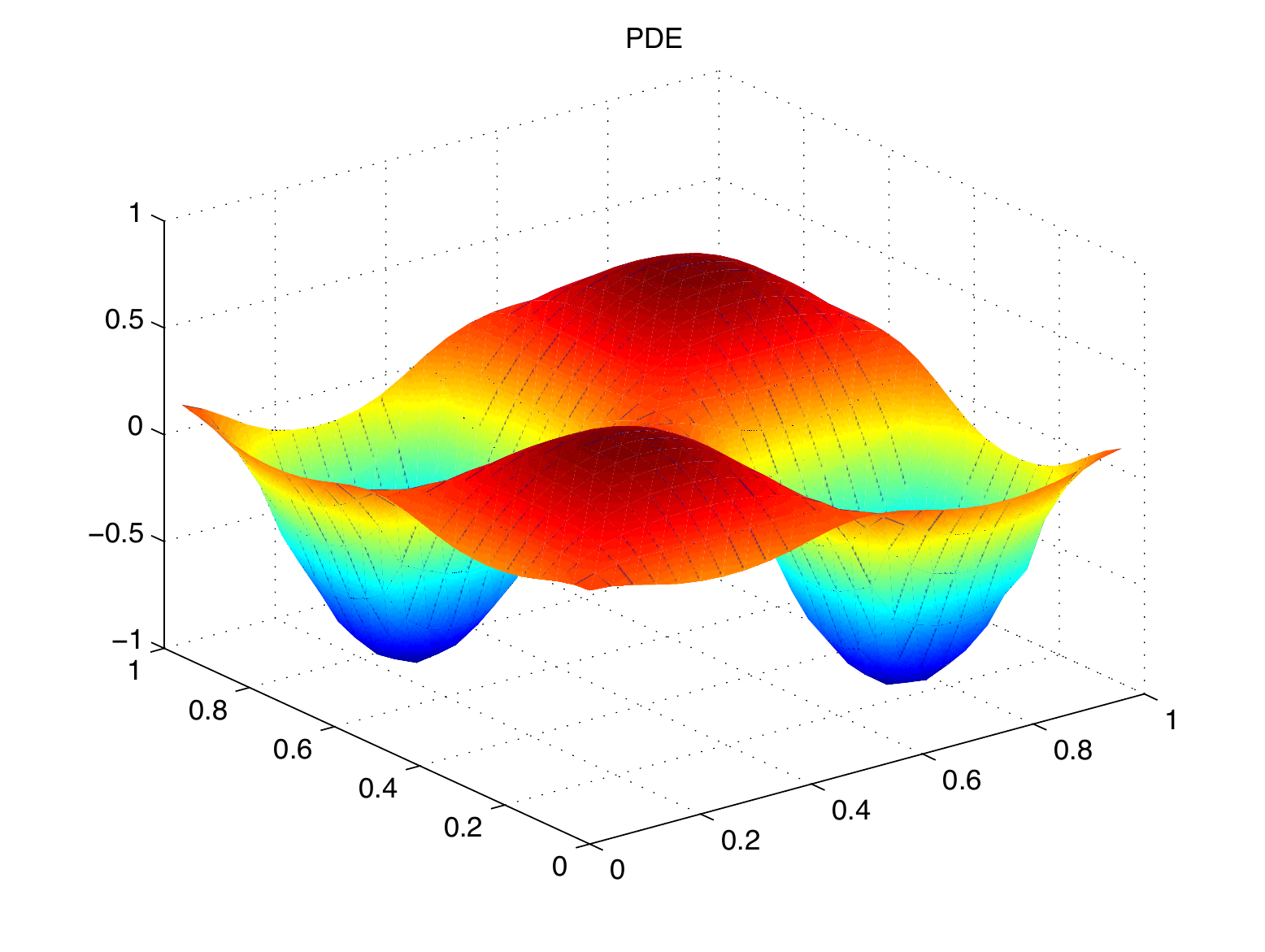} \\
\includegraphics[width=3.0in]{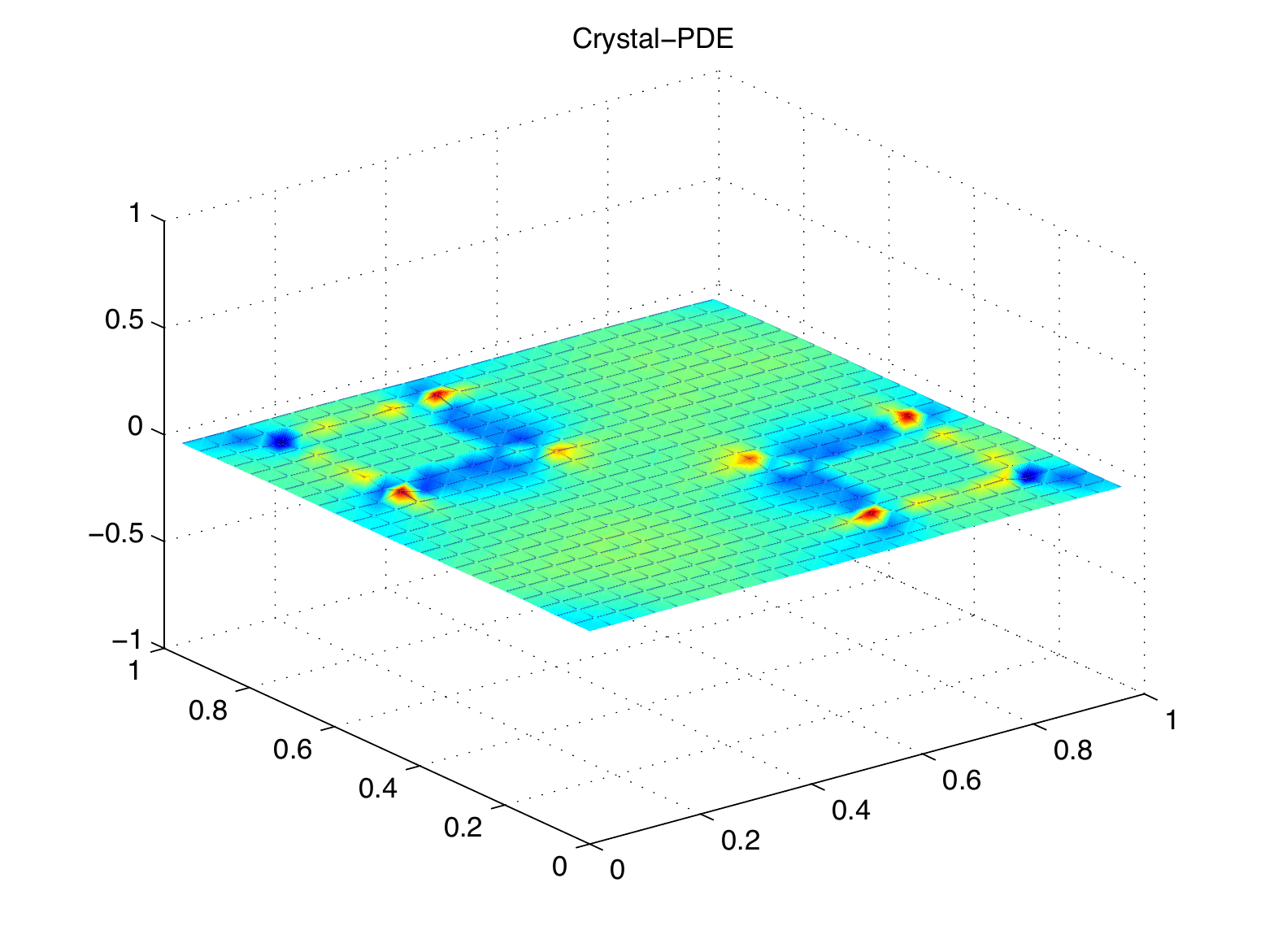}
\caption{The microscopic profile in the rough scaling (top), the solution of PDE \eqref{roughpde} (middle),  and the difference between the two (bottom) in 2+1 dimensions for $K=1.5$ with $V(z)=z^2$ at $T=10^{-30}.$ 
 The maximum of the difference between the rescaled microscopic profile and the PDE solution is roughly $8\times 10^{-3}.$}
\label{fig:p2rough2d}
\end{figure}

\begin{figure}
\centering
\includegraphics[width=3.0in]{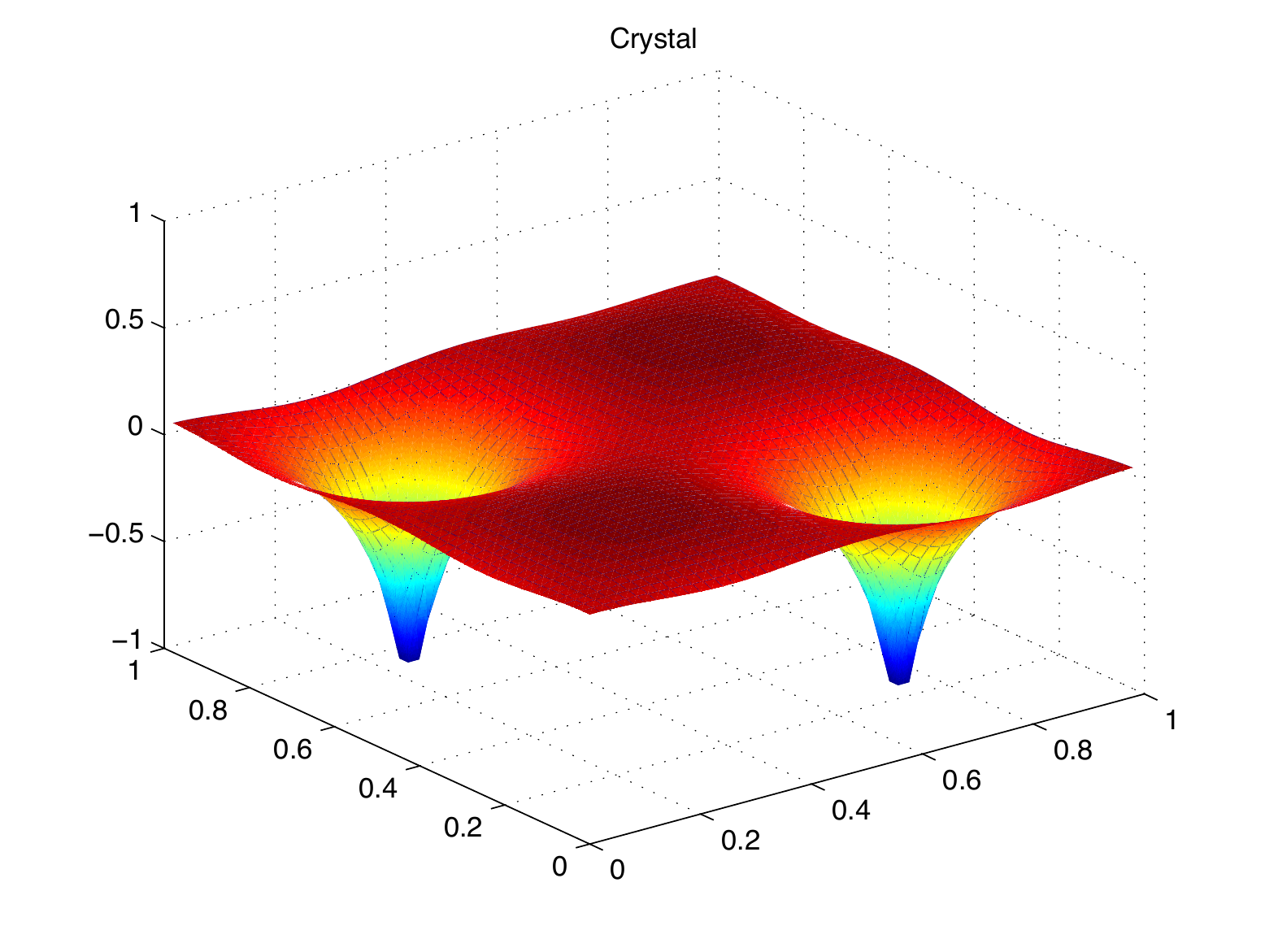}
\includegraphics[width=3.0in]{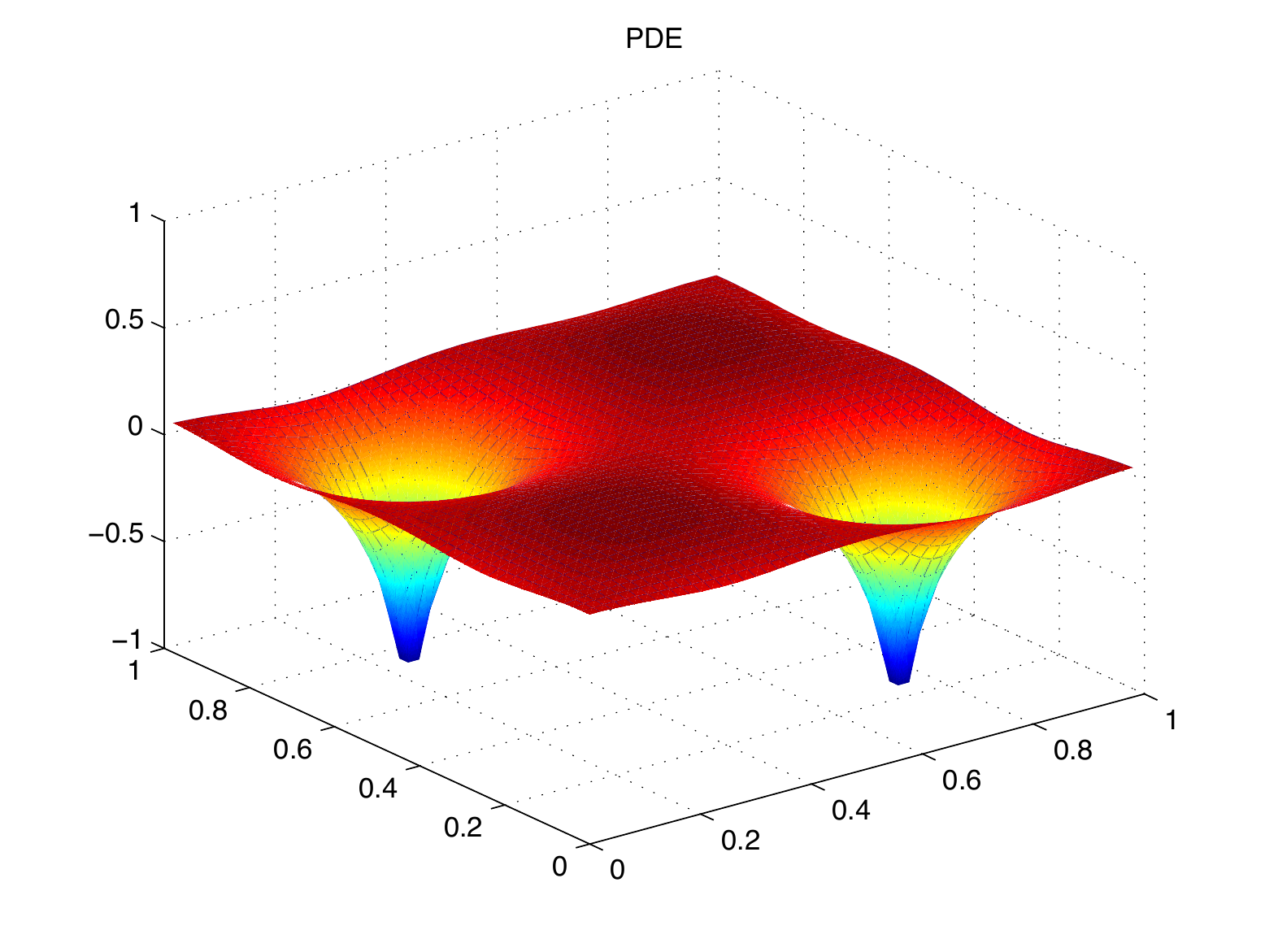}
\includegraphics[width=3.0in]{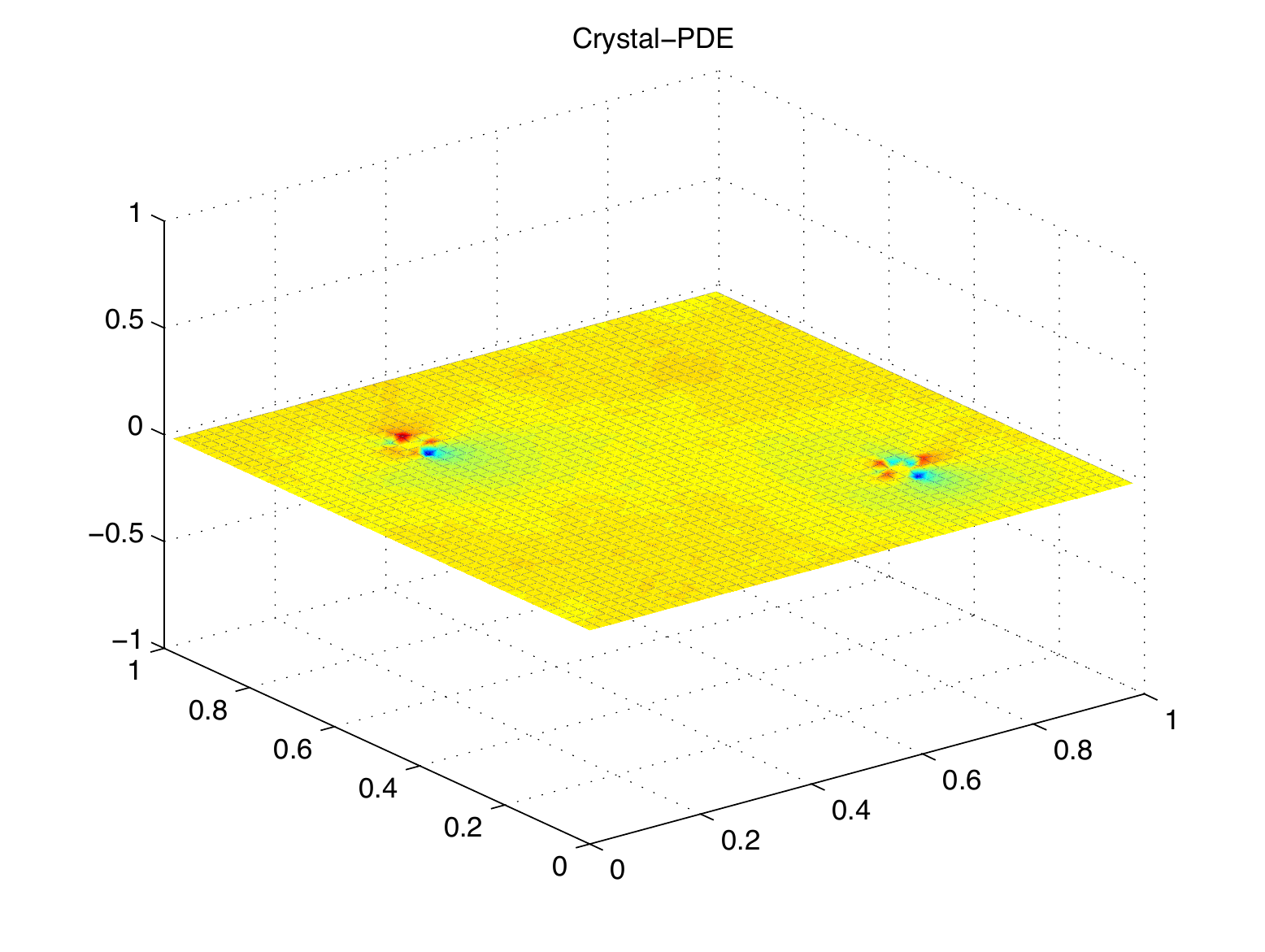}
\caption{The microscopic profile in the rough scaling (top), the solution of PDE \eqref{roughpde} (middle), and the difference between the two (bottom) in 2+1 dimensions for $K=1.5$ with  $V(z)=z^2$ at $T=10^{-10}$.  Note the formation of cusp-like solutions.}
\label{fig:p2rough2d-cusp}
\end{figure}

Having investigated the convergence of the rescaled microscopic evolutions in both scaling regimes in $1+1$ dimensions it is natural to ask if our conclusions are also valid in 2+1 dimensions.  In short, the answer seems to be yes.  In fact, our results in 2+1 dimensions are exactly analogous to those in $1+1$ dimensions.  In 
Figures \ref{fig:p1smooth2d} and \ref{fig:p2smooth2d} we find that, for $V(z)=|z|$ and $V(z)=z^2,$  the agreement between the rescaled  microscopic evolution and the PDE \eqref{smoothpde} in the smooth scaling limit is compelling.  Figure \ref{fig:p2rough2d} presents similar results in the $V(z)=z^2$ case for the rough scaling limit.  As in the $1+1$ dimensional case, in $2+1$ dimensions, the evolution in the rough scaling limit seems to form singularities in convex regions of the surface (see Figure \ref{fig:p2rough2d-cusp}).

\begin{figure}
\centering
\includegraphics[width=2.5in]{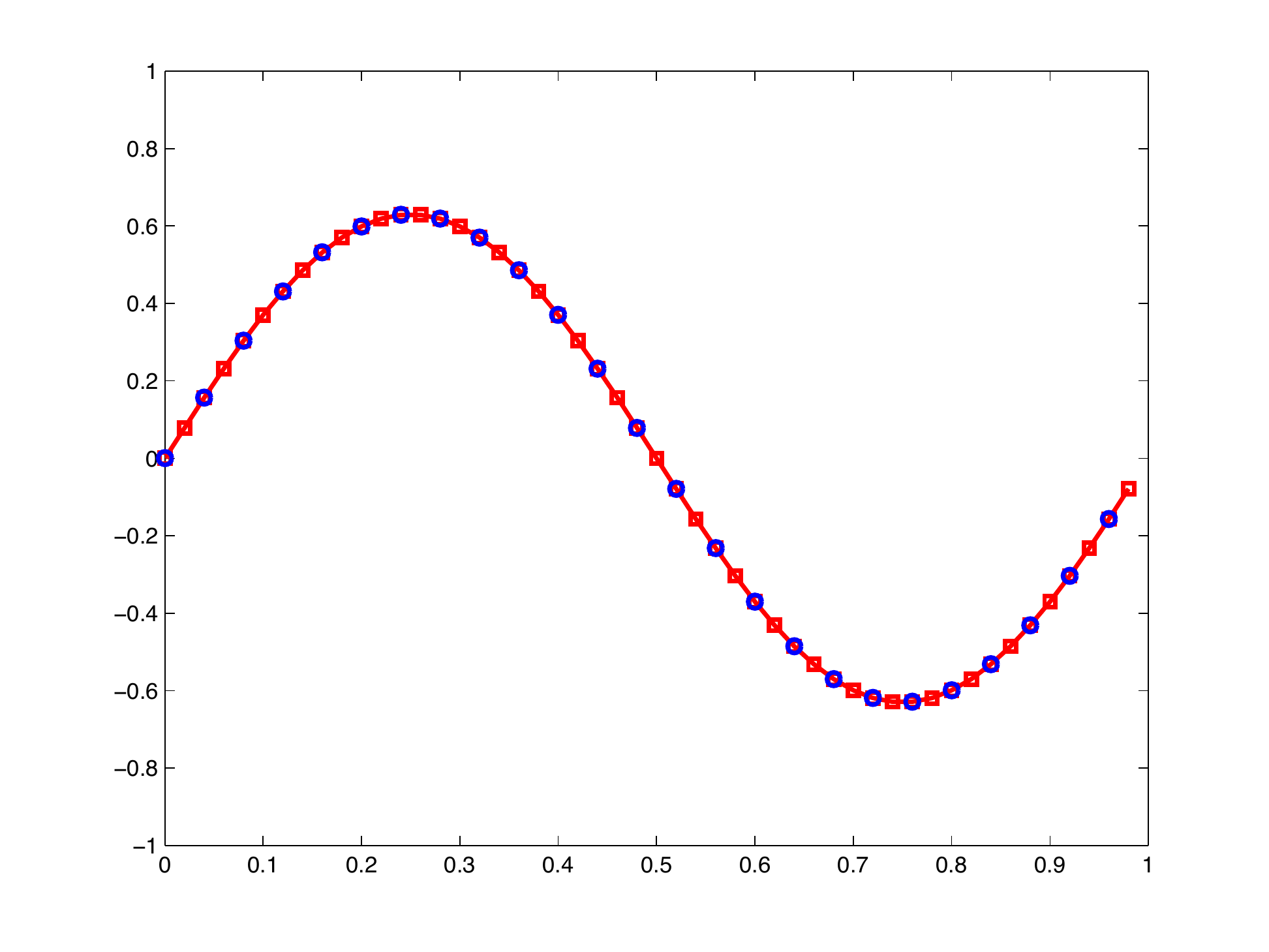}
\includegraphics[width=2.5in]{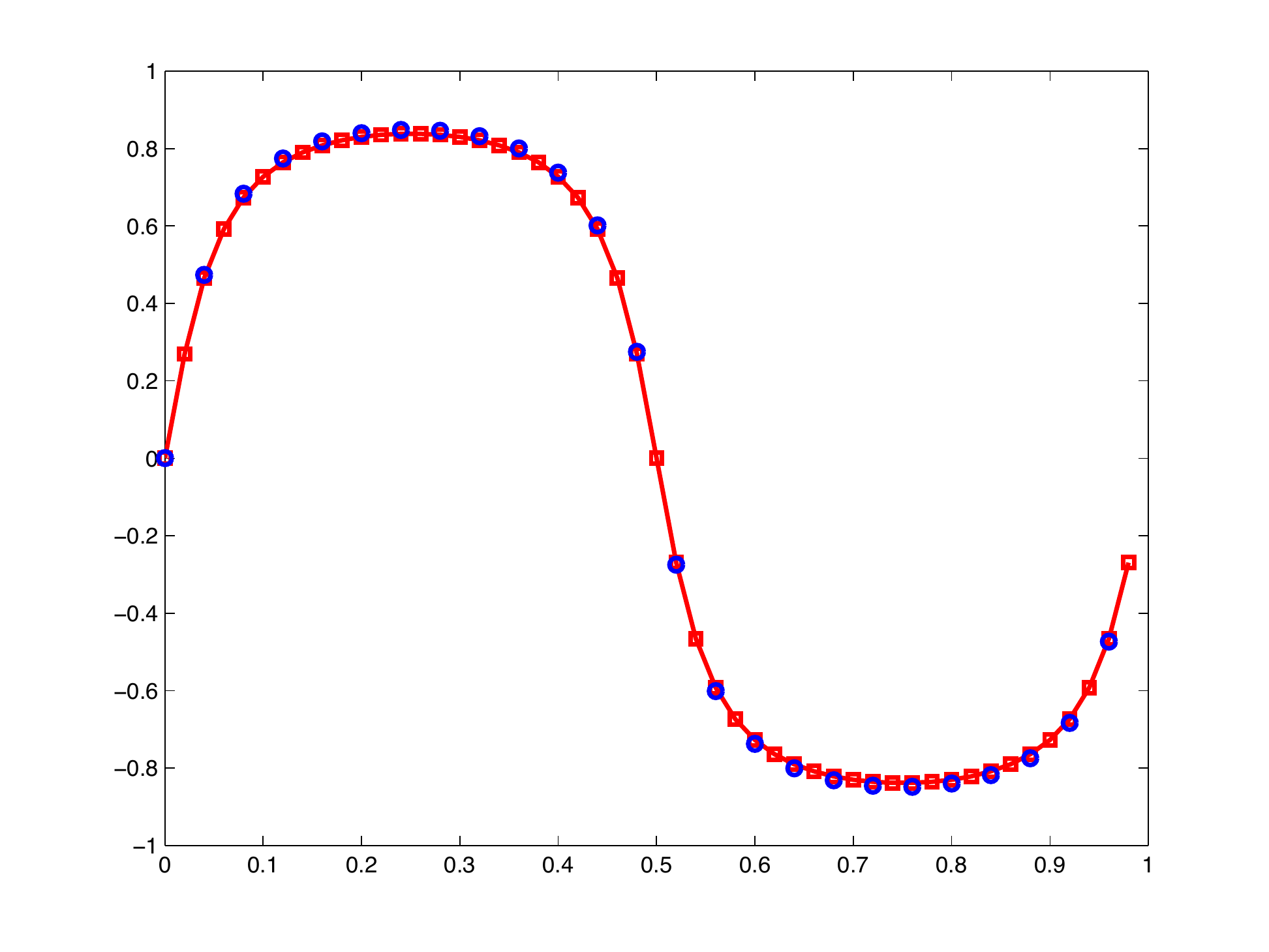}
\includegraphics[width=2.5in]{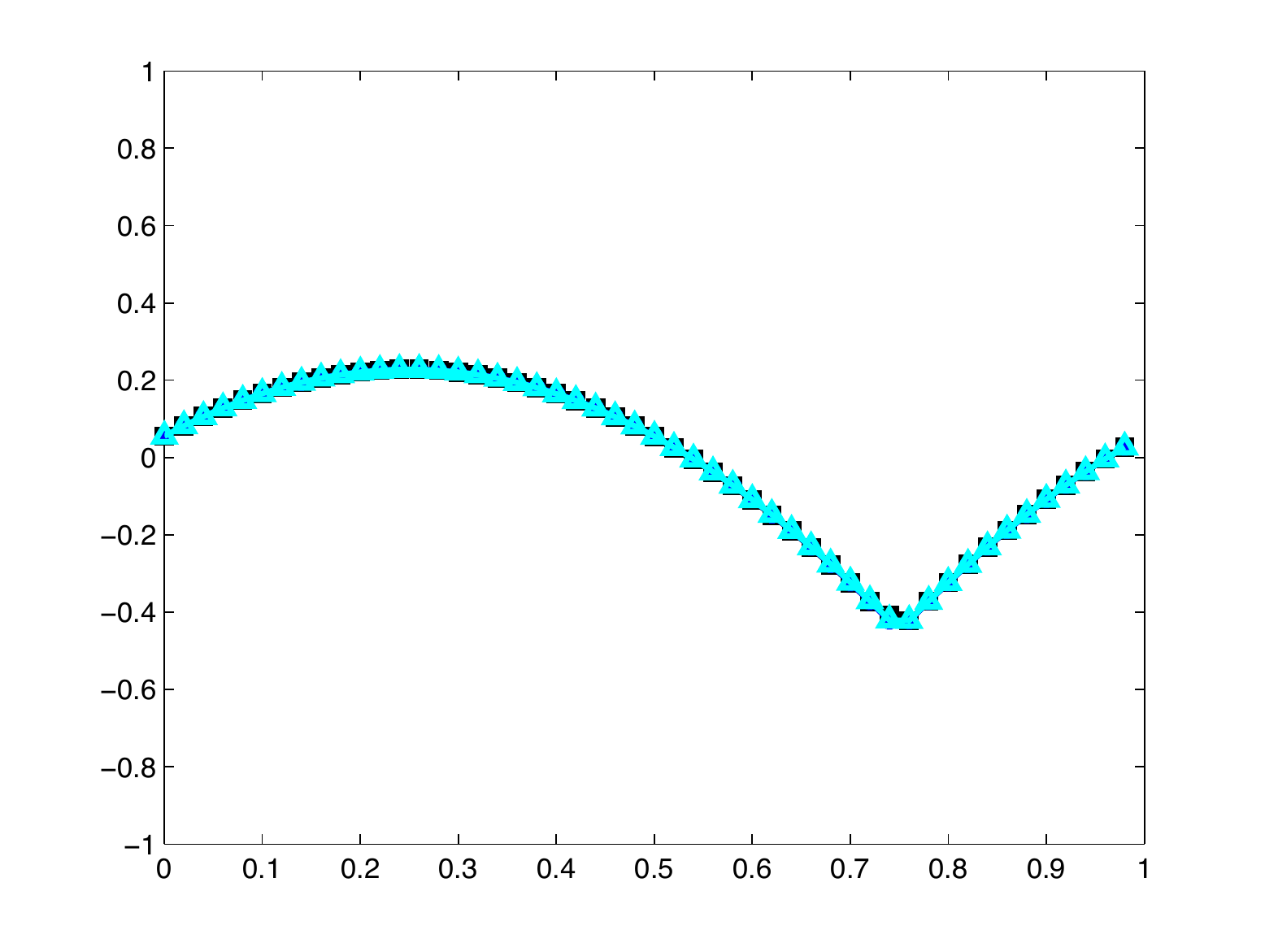}
\caption{Results of fix-point iteration in which  PDE \eqref{smoothpde} (top left and right) and \eqref{roughpde} (bottom) in 1+1 dimensions are evolved for some interval of time, then rescaled to have maximum height (in absolute value) equal to 1 and then evolved and rescaled repeatedly until convergence.   The solutions appear to be approximately of the form $h(t,x) = g(x) \phi (t).$  The plots depict the function $g$ corresponding to each PDE.   Equation \eqref{smoothpde} was evolved for intervals of length $T = 2\time 10^{-4}$  for $V(z)=z^2$ (top left) and $T = 10^{-3}$ for $V(z)=|z|$ (top right).  For the rough crystal, we take the intervals of size $T=10^{-10}$ for $V(z)=z^2$ (bottom).}
\label{fig:selfsimilar}
\end{figure}

There are many interesting features of the behavior of the microscopic system in these two scaling regimes left to explore.  For example, as we have already remarked,  the rough scaling regime seems to produce cusps in convex regions while the smooth scaling regime seems to have a smoothing effect on non-smooth surfaces.  Below we offer very preliminary numerical evidence suggesting a few additional interesting questions about the qualitative behavior of the microscopic system at large scales.

One might ask about the behavior of the surfaces as they near equilibrium ($h\equiv 0$).  In Figure \ref{fig:selfsimilar} we show that the surfaces appear to approximately factor as $h(t,x) = \phi(t) g(x)$ for very large $t.$  The results in that figure were generated via a fixed point iteration in which the surface is evolved for some length of time and then rescaled so that the surface's maximal (in absolute value) height is 1, and then evolved and rescaled again and so on.  Each plot shows the last two iterations of that fixed point iteration (before rescaling).  The overlap in those surfaces indicates that the iteration has converged (to $g(x)$).    We note that the function $g(x)$ will typically have some dependence on the particular initial profile.  As above we used $\sin(2\pi x)$ in these simulations.

\begin{figure}
\centering
\includegraphics[width=4in]{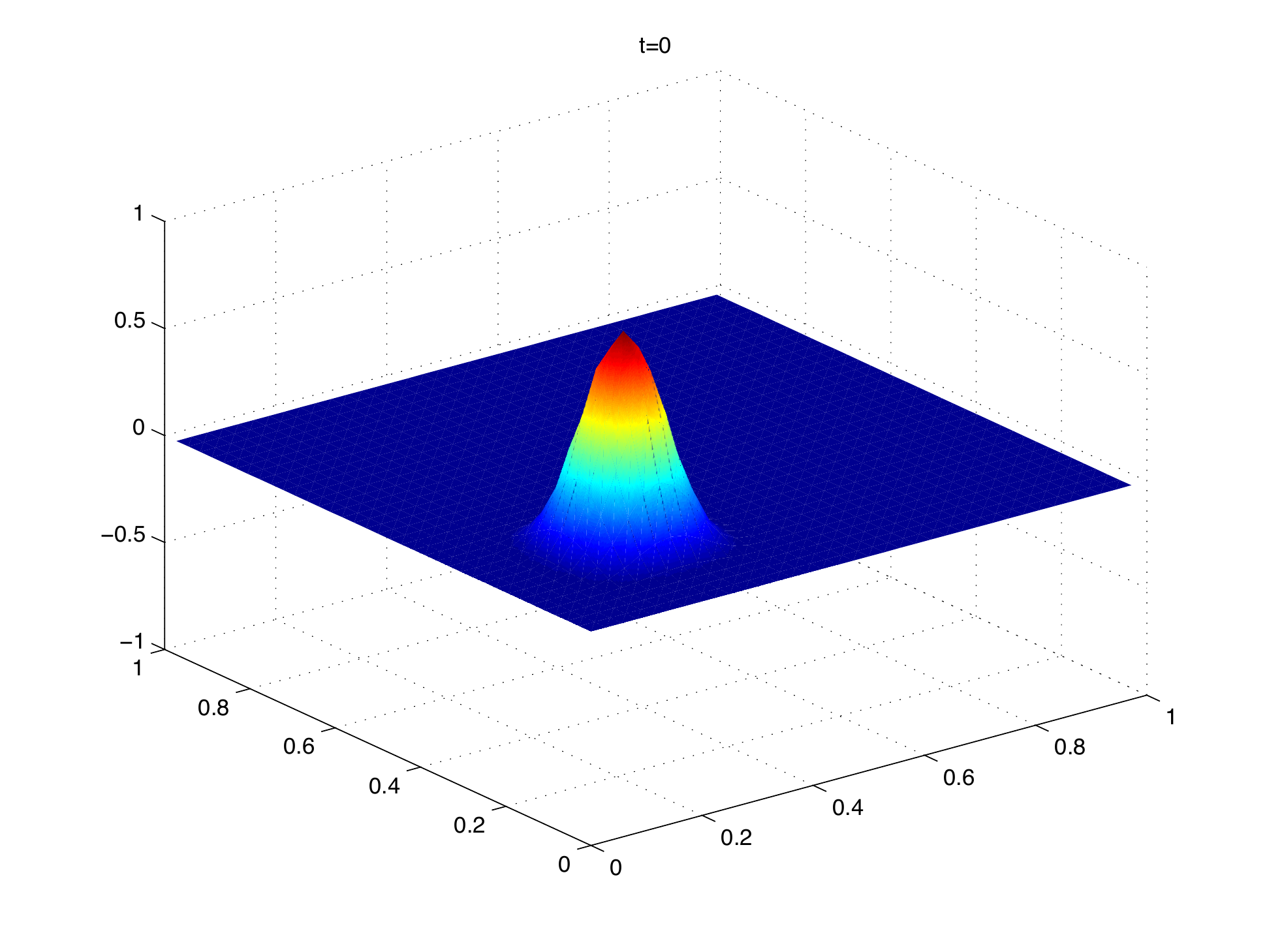}
\caption{Surface plot of  initial profile \eqref{eqn:initdata2} used in wetting experiments.  The profile is non-zero only in the lower left quadrant of the domain.}
\label{fig:wetting_id}
\end{figure}

\begin{figure}
\centering
\includegraphics[width=2.5in]{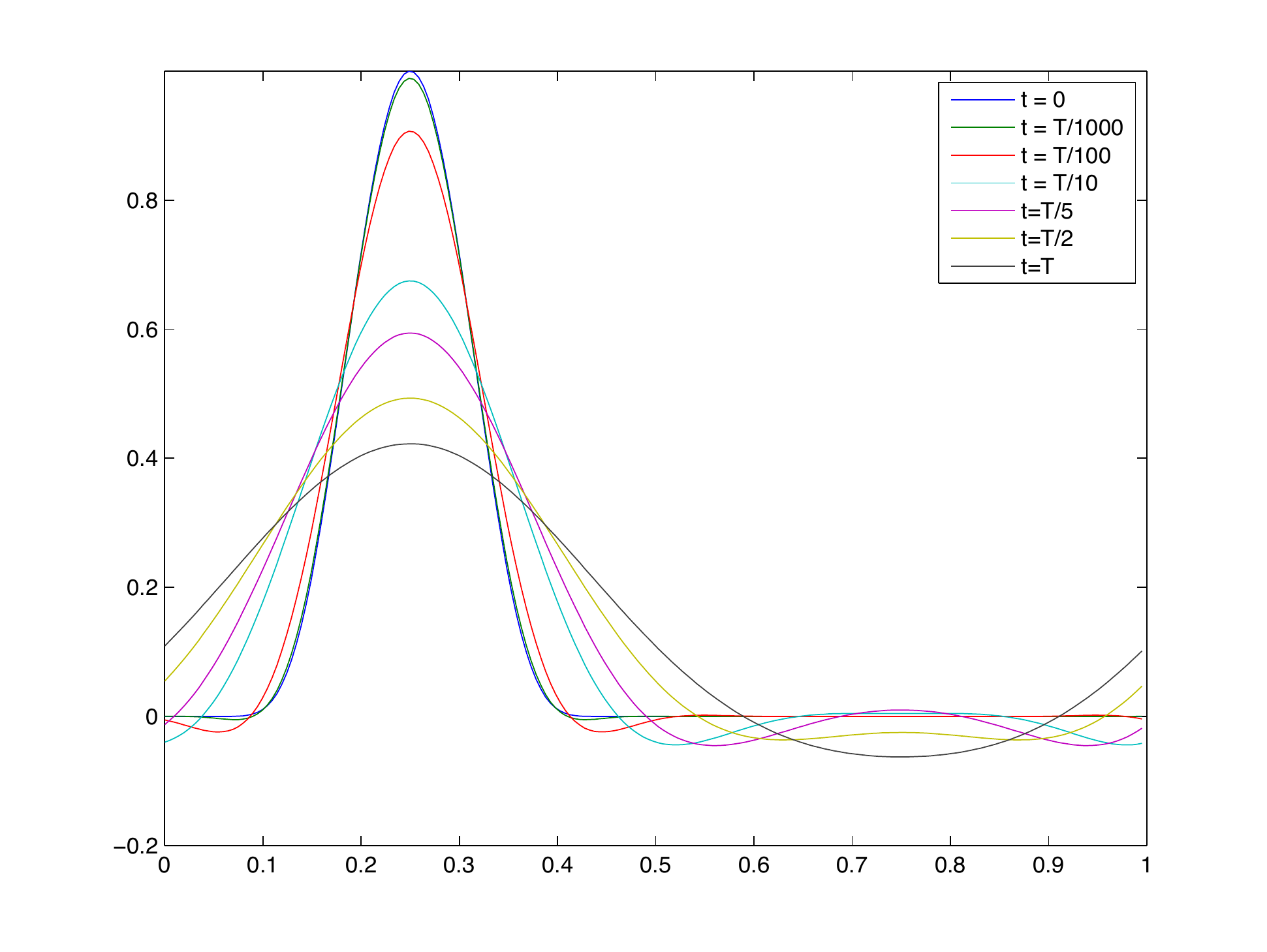}
\includegraphics[width=2.5in]{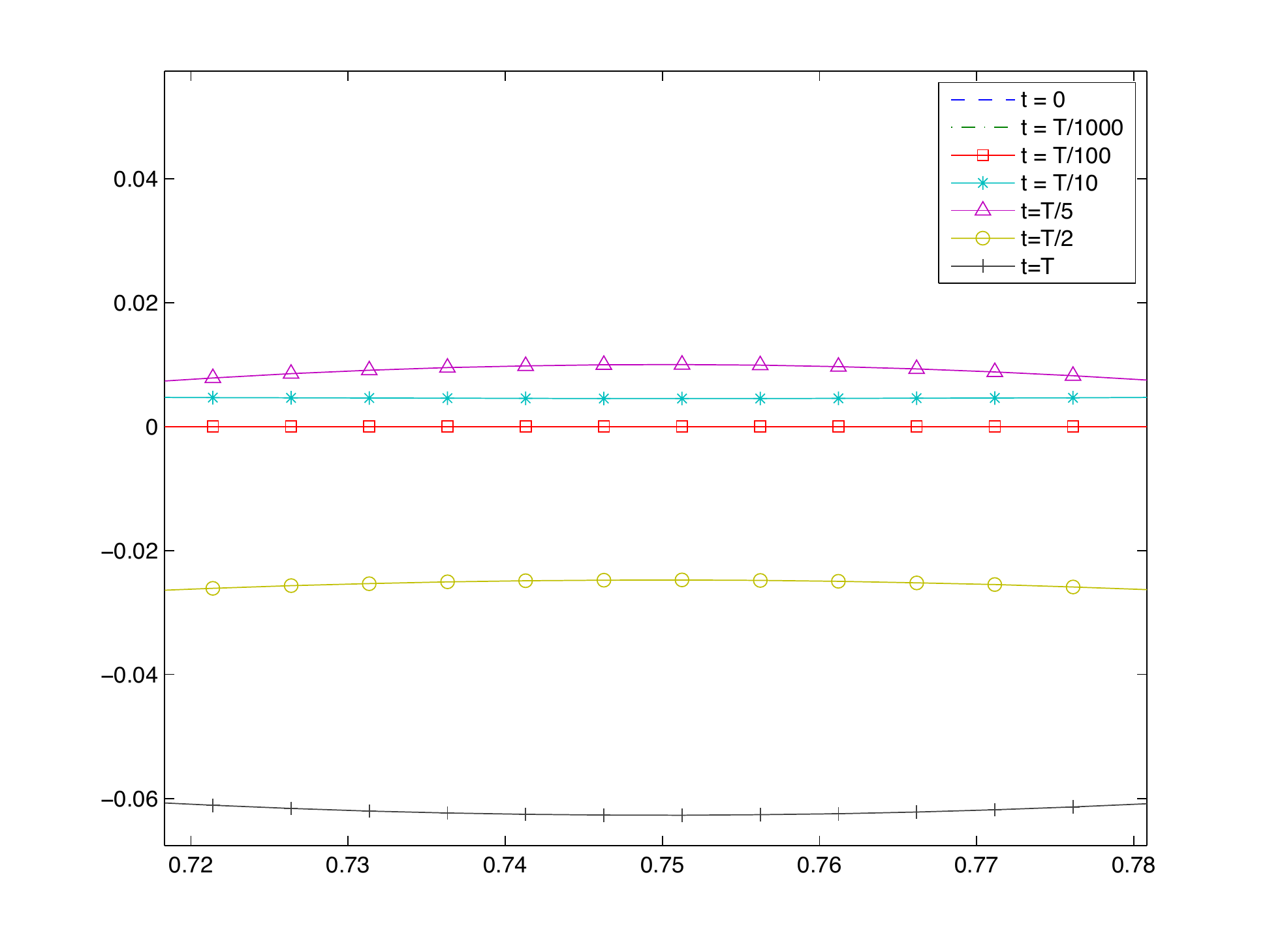}
\caption{Snapshots of solution of PDE \eqref{smoothpde} in 1+1 dimensions with $V(z)=z^2,$ $K=1.5,$ from the initial profile in \eqref{eqn:initdata2}, at times in an interval of length $T=5\times 10^{-5}$ (left) and a blowup in the region of zero initial height (right).}
\label{fig:wetting1}
\end{figure}

\begin{figure}
\centering
\includegraphics[width=2.5in,height =1.88in]{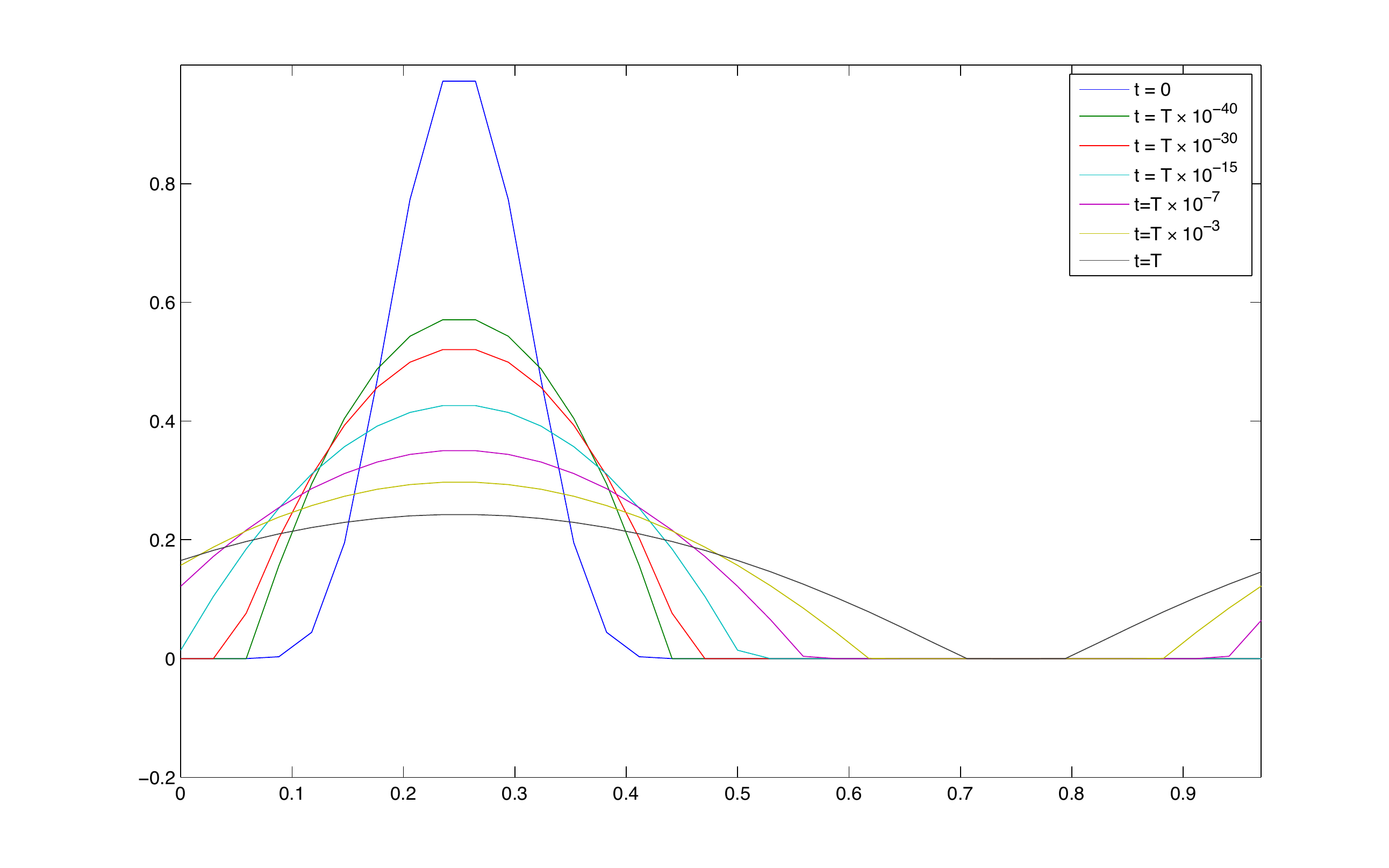}
\includegraphics[width=2.5in]{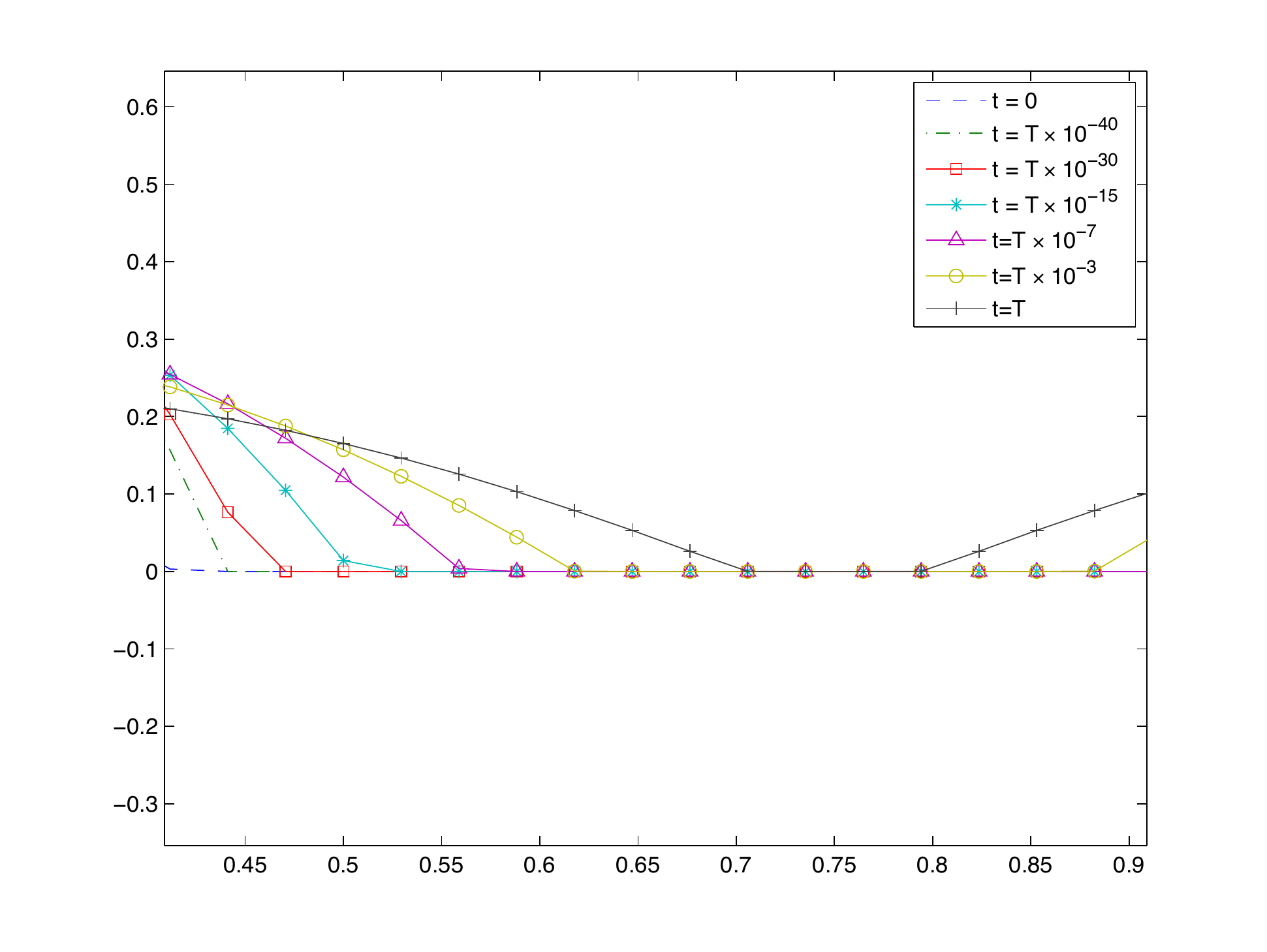}
\caption{Snapshots of solution of PDE \eqref{roughpde} in 1+1 dimensions with $V(z)=z^2,$ $K=1.5,$  from the initial profile in \eqref{eqn:initdata2}, at times in an interval of length $T=5\times 10^{-7}$ (left) and a blowup in the region of zero initial height (right).}
\label{fig:wetting3}
\end{figure}

\begin{figure}
\centering
\includegraphics[width=2.5in]{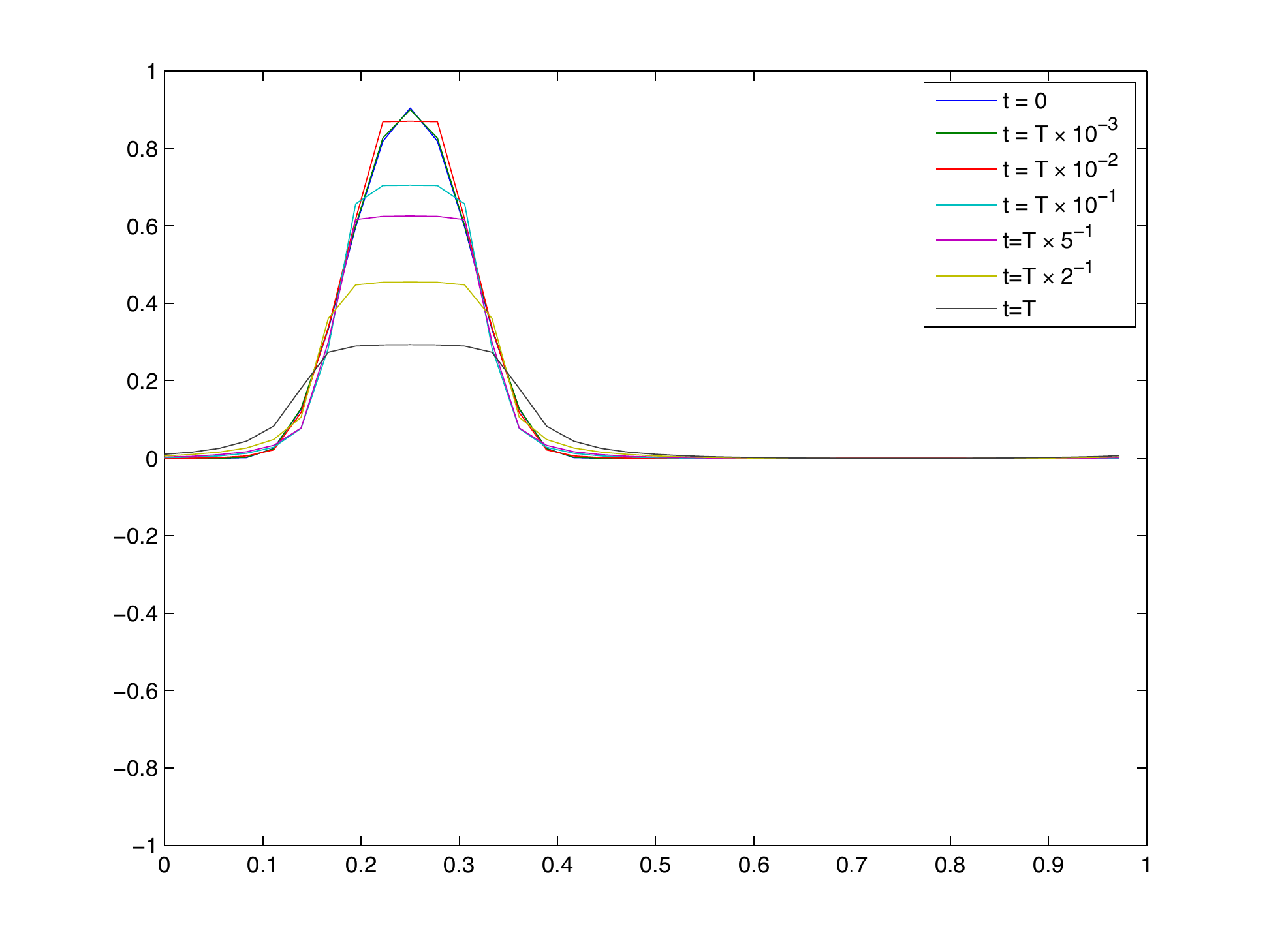}
\includegraphics[width=2.5in]{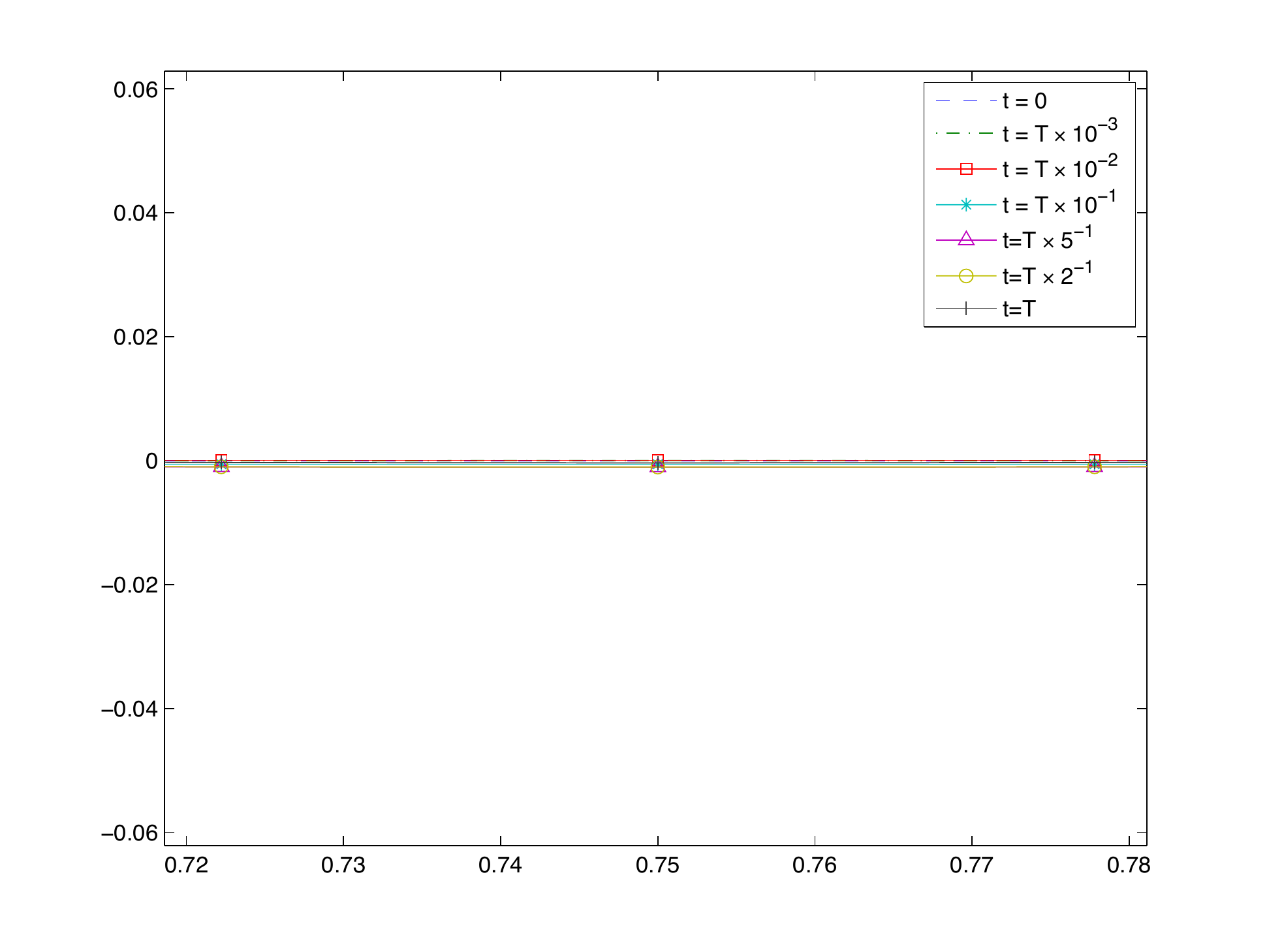}
\caption{Snapshots of 1 dimensional cross section at $x=0.25$ of solution of PDE \eqref{smoothpde} in 2+1 dimensions with $V(z)=|z|,$ $K=5,$  from the initial profile in \eqref{eqn:initdata2}, at times in an interval of length $T=4\times 10^{-5}$ (left) and a blowup in the region of zero initial height (right).
 }
\label{fig:wetting2_2d}
\end{figure}

Another interesting feature of these scaling limits to explore is the possibility the rate at which regions of non-zero height spread into regions of zero height.  We will refer to this process as wetting.  In order for facets (macroscopic flat regions on the crystal surface) to be stable features of a surface, the wetting rate should be finite.    Given the preliminary  tests reported in Figure \ref{fig:wetting1} it seems suggestive that, in the 1+1 case, the smooth PDE \eqref{smoothpde} (at all temperatures and for both $V(z)=|z|$ and $V(z)=z^2$) wets infinitely quickly.  At least in the $V(z)=z^2$ case, this is as one might expect for a PDE that is similar to the fourth order heat equation $\partial_t h = - K \partial_x^4 h.$  As reported in Figure \ref{fig:wetting3}, the rough PDE \eqref{roughpde} in both 1+1 and 2+1 dimensions seems to wet at a finite rate.  It also seems possible that the smooth PDE in 2+1 dimensions with $V=|z|$ can wet at finite rate at least for large enough temperature (see Figure \ref{fig:wetting2_2d}), though our evidence for this is weak.  In both 1+1 dimensions and 2+1 dimensions the wetting rate was investigated for an initial profile of the form
\begin{equation}
\label{eqn:initdata2}
h(0,x) = \left\{ \begin{array}{c}
 e^{8-|x|^{-1} - (0.5-|x|)^{-1}} \ \  \text{for} \ 0 < |x| < \frac12, \\
0 \ \ \text{otherwise},
\end{array} \right.
\end{equation}
This initial profile in $2$ dimensions is plotted in Figure \ref{fig:wetting_id}.

\begin{figure}
\centering
\includegraphics[width=2.5in]{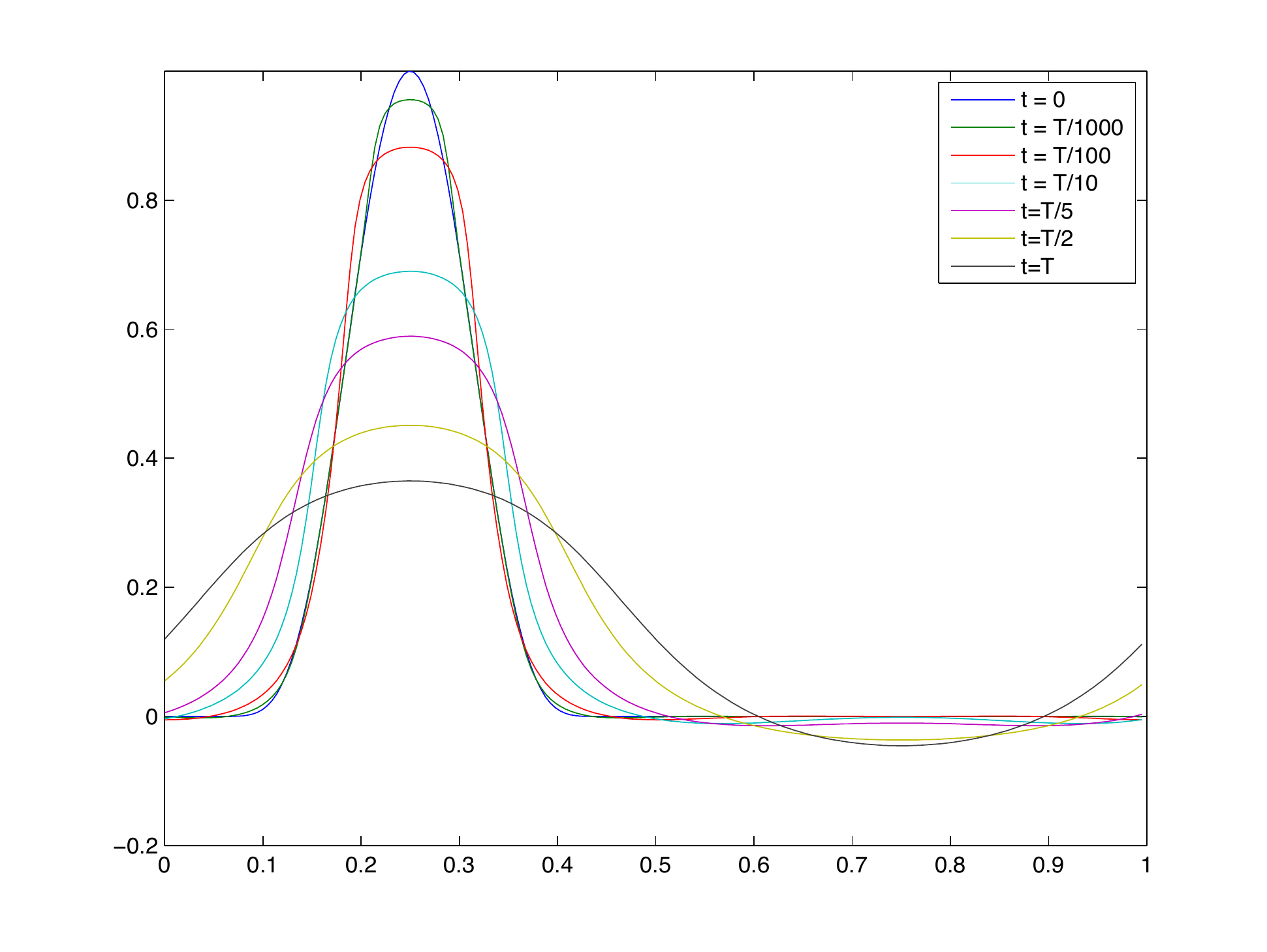}
\includegraphics[width=2.5in]{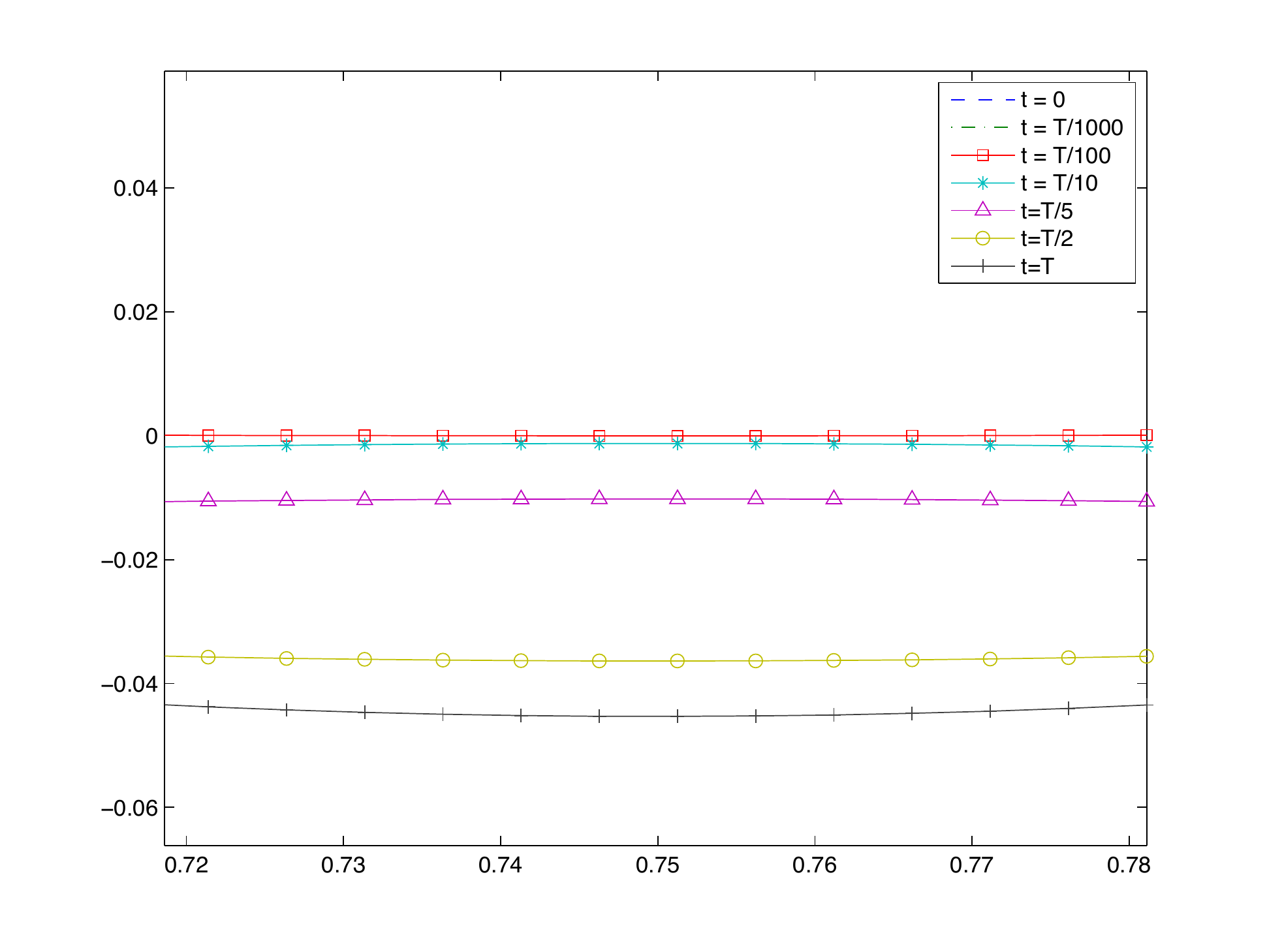}
\caption{Snapshots of solution of PDE \eqref{smoothpde} in 1+1 dimensions with $V(z)=|z|,$ $K=1.5,$  from the initial profile in \eqref{eqn:initdata2}, at times in an interval of length $T=5\times 10^{-4}$ (left) and a blowup in the region of zero initial height (right).}
\label{fig:wetting2}
\end{figure}

\begin{figure}
\centering
\includegraphics[width=2.5in]{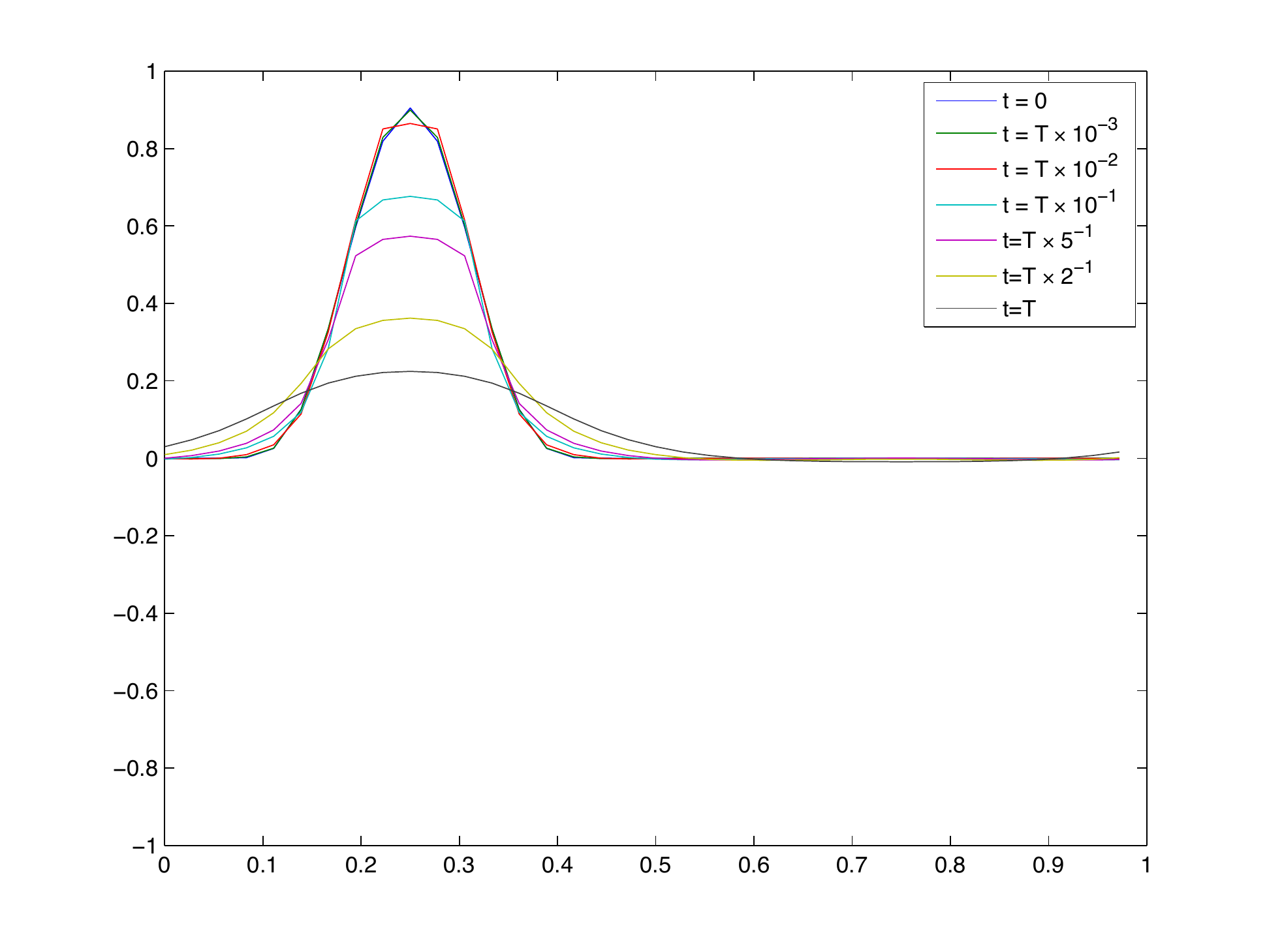}
\includegraphics[width=2.5in]{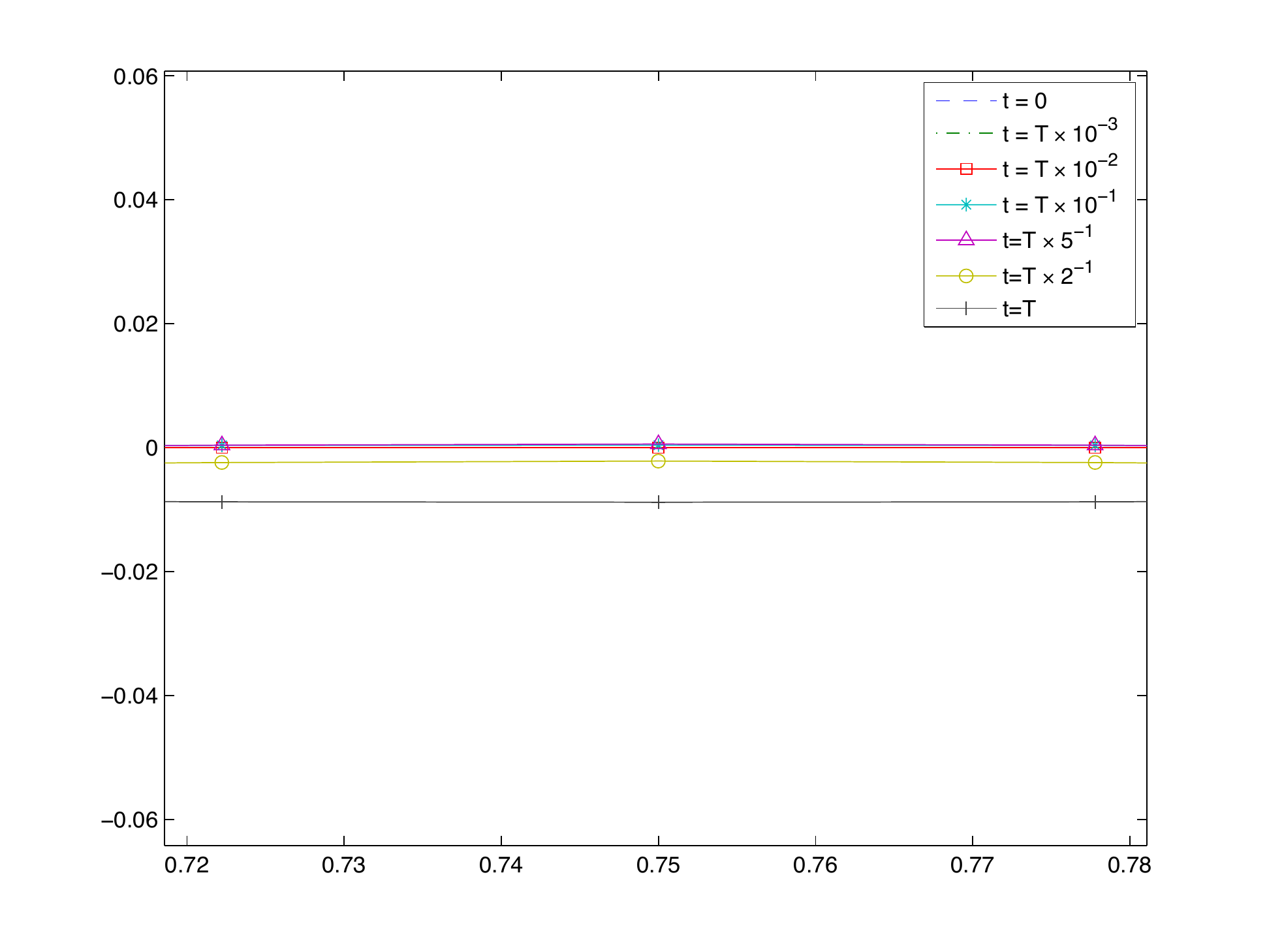}
\caption{Snapshots of 1 dimensional cross section at $x=0.25$ of solution of PDE \eqref{smoothpde} in 2+1 dimensions with $V(z)=|z|,$ $K=1.5,$  from the initial profile in \eqref{eqn:initdata2}, at times in an interval of length $T=2\times 10^{-4}$ (left) and a blowup in the region of zero initial height (right).
}
\label{fig:wetting1_2d}
\end{figure}

We caution that these qualitative features are difficult to conclusively determine numerically and our tests are only meant to be suggestive.    Only rigorous mathematical analysis can answer these questions definitively.
Given the strong agreement demonstrated here between the rescaled microscopic model and equations \eqref{smoothpde} and \eqref{roughpde}, it seems safe to pursue these and other questions about the large scale qualitative behavior of the microscopic model at the level of the PDE.  In future work we will pursue these questions along with rigorous mathematical proofs of the convergence claims in this paper.

\section{Informal derivations of the PDE limits}
\label{sec:PDEder}

In this section we offer further evidence in support of our PDE limits in the form of informal derivations.  These derivations are not rigorous but offer insight into why the PDE \eqref{smoothpde} and \eqref{roughpde} arise.  The arguments follow a standard line of reasoning in the literature on hydrodynamic limits (see e.g. \cite{GPV,Funaki}).

\subsection{The smooth scaling regime}
\label{s:smoothkrug}

Consider sums of $h_N(t,\cdot)$ against the sampled values of some smooth, periodic  function  $v$ on $[0,1]$ i.e.
quantities of the form
\[
\varphi_N(t) = N^{-(1+d)} \sum_{\alpha\in \mathbb{T}^d_N} h_N(N^4 t,\alpha) v(N^{-1}\alpha).
\]
Appealing to the smoothness of $v$ we have that
\[
\varphi_N(t) \approx \int \bar h_N(t,x) v(x) dx
\]
where in this subsection the overbar represents the projection defined in \eqref{smoothscaling}.
  Notice that 
  \begin{equation}\label{gen2}
\mathcal{A}_N \left[ \sum_{\alpha\in \mathbb{T}^d_N} h_N(\alpha) v(N^{-1}\alpha) \right]  = - \sum_{\substack{\beta\in\mathbb{T}^d_N \\ |\beta-\alpha |=1}} ( r_N(\beta) - r_N(\alpha)).
\end{equation}
Therefore we can write
 \begin{align*}
& \varphi_N(t) - \varphi_N(0)
  =   \\
  & \hspace{.5cm} -N^{3-d} \int_0^t \sum_{\substack{\beta\in\mathbb{T}^d_N \\ |\beta-\alpha |=1}} (r_N(N^4 s,\beta) - r_N(N^4 s,\alpha))v(N^{-1}\alpha) ds + M_\varphi(t),
  \end{align*}
  where, for each $t$ and $\alpha,$
  $\mathbf{E}\left[M_\varphi(t)\right]=0.$  We expect the last term to vanish as $N\rightarrow \infty$ so we will drop it in the following.
  
    Summing by parts, this formula can be re-expressed as
 \begin{align*}
& \varphi_N(t) - \varphi_N(0)
  \approx \\
  & \hspace{.5cm} {N^{3-d}} \int_0^t \sum_{\beta\in \mathbb{T}^d_N }  r_N(N^4 s,\beta)  \sum_{\substack{\alpha\in\mathbb{T}^d_N \\ |\beta-\alpha |=1}} v\left( N^{-1}\alpha \right)- v\left(N^{-1}{\beta}\right) ds,
    \end{align*}
    which, appealing to the smoothness of $v,$ is approximated by
     \begin{equation*}
\varphi_N(t) - \varphi_N(0) 
  \approx N^{1-d} \int_0^t  \sum_{\beta\in \mathbb{T}^d_N }  r_N(N^4 s,\beta)  \Delta  v \left( N^{-1}\beta\right) ds.
    \end{equation*}
    
    At this point we assume  that the random variables $r_N(s,\beta)$ {\it locally equilibrate} on a time scale much faster than $\mathcal{O}(N^4)$ to their equilibrium (long time) distribution conditioned on the profile $\bar h_N(s,\cdot).$  This conditional equilibrium distribution is the one implied by the equilibrium distribution $\rho^m_N$ for $h_N.$  The resulting, locally equilibrated random variables $\tilde r_N(N^4s, \beta)$ are being summed in the last display against a smooth function.  Therefore we can expect that a Law of Large Numbers applies and the locally equilibrated random variables $\tilde r_N$ in this expression can be replaced by their expectations, i.e. by there expectations under the conditional equilibrium distribution.  If we also assume that any dependence between the $\tilde r_N$ is negligible for large $N,$ then a version of the conditional limit theorems (see e.g. \cite{GPV}) implies that in the large $N$  limit the conditional equilibrium distribution is well approximated by the so called optimal exponential twist
    \begin{align*}
 &   \rho^{m,\sigma_D}_N(\nabla^+ \tilde h_N)  = \\
 & \hspace{.5cm} \frac{1}{\mathcal{Z}^{m,\sigma_D}_N} \exp\left({-K \sum_{\substack{\alpha\in \mathbb{R}^d_N\\ i\leq d}} V(\nabla^+_i \tilde h_N(\alpha))+ K \sigma_D(\nabla^+ h_N(\alpha))^\text{\tiny T} \nabla^+ \tilde h_N(\alpha)}\right)
    \end{align*}
    where $\sigma_D$ was defined above in \eqref{eqn:sigmaD} and $\mathcal{Z}^{m,\sigma_D}_N$ is a normalization constant.  Note that $h_N$ should be regarded as a fixed (non-random) parameter in this expression. 
    
    Fortunately, the expectation of the rates $\tilde r_N$ under $\rho^{m,\sigma_D}_N$ takes a very simple form.
    To see this, first notice that our generalized coordination number satisfies the relation
\begin{equation}\label{biggyz}
2 n_N(\alpha) +  \sum_{\substack{\beta\in\mathbb{T}^d_N\\ i\leq d} } V(\nabla V^+_i h_N (\beta)) =
 \sum_{\substack{\beta\in\mathbb{T}^d_N\\ i\leq d} } V(\nabla^+_i J_\alpha h_N (\beta)).
\end{equation}
This implies that
\begin{align*}
& \big\langle \tilde r_N(\alpha) \big\rangle^{m,\sigma_D}_N = \frac{1}{2d} \sum_{\tilde h_N}  e^{-2 K \tilde n_N(\alpha)} \rho^{m,\sigma_D}_N(\nabla^+ \tilde h_N) \\
& \hspace{1cm} = \frac{1}{2d\,\mathcal{Z}^{m,\sigma_D}_N}\sum_{\tilde h_N} e^{-K 2 \tilde n_N(\alpha) -K \sum_{\substack{\beta\in\mathbb{T}^d_N\\ i\leq d} } V(\nabla V^+_i \tilde h_N (\beta)) }\\
&\hspace{2cm} \times  e^{  K \sum_{\beta \in \mathbb{T}^d_N} \sigma_D(\nabla^+ h_N(\beta))^\text{\tiny T} \nabla^+ \tilde h_N(\beta) }\\
&  \hspace{1cm} = \frac{1}{2d \mathcal{Z}^{m,\sigma_D}_N}\sum_{\tilde h_N} e^{-K \sum_{\beta\in\mathbb{T}^d_N}V(\nabla^+_i J_\alpha \tilde h_N (\beta)) }
 e^{K\sum_{\beta\in\mathbb{T}^d_N}   \sigma_D(\nabla^+ h_N(\beta))^\text{\tiny T} \nabla^+ J_\alpha \tilde h_N(\beta)}
 \\
 &\hspace{2cm} \times 
 e^{-K\sum_{i\leq d} \sigma_{D,i} (\nabla^+ h_N(\alpha))-\sigma_{D,i}(\nabla^+ h_N(\alpha-e_i))}
\end{align*}
In this last equation we can carry out the summation over $J_\alpha \tilde h_N$ instead of $\tilde h_N$ to obtain
\[
\big\langle \tilde r_N(\alpha) \big\rangle^{m,\sigma_D}_N = \frac{1}{2d}e^{ -K\sum_{i\leq d} \sigma_{D,i} (\nabla^+ h_N(\alpha))-\sigma_{D,i}(\nabla^+ h_N(\alpha-e_i)) }\, \frac{\mathcal{Z}^{m-1,\sigma_D}_N}{2d \mathcal{Z}^{m,\sigma_D}_N}.
 \]

Ignoring the factor of $\mathcal{Z}^{m-1,\sigma_D}_N / \mathcal{Z}^{m,\sigma_D}_N$ which will be small for large $N,$ and summarizing the above discussion we arrive at the expression
     \begin{multline*}
\varphi_N(t) - \varphi_N(0) 
  \approx N^{1-d} \int_0^t  \sum_{\beta\in \mathbb{T}^d_N } (2d)^{-1} \Delta  v \left( N^{-1}\beta\right) \\
  \times e^{ -K\sum_{i\leq d} \sigma_{D,i} (\nabla^+ h_N(N^4 s, \beta))-\sigma_{D,i}(\nabla^+ h_N(N^4 s, \beta-e_i)) } ds.
    \end{multline*}

   We can rewrite the right hand side of the last display in terms of $\bar h_N$ to obtain
        \begin{multline*}
\varphi_N(t) - \varphi_N(0) 
  \approx N^{1-d} \int_0^t  \sum_{\beta\in \mathbb{T}^d_N }(2d)^{-1} \Delta  v \left( N^{-1}\beta\right) \\
  \times e^{ -K\sum_{i\leq d} \sigma_{D,i} (N \nabla^+ \bar h_N( s, N^{-1}\beta))-\sigma_{D,i}(N \nabla^+ \bar h_N(s, N^{-1}(\beta-e_i))) } ds.
    \end{multline*}
  Here we have abused notation slightly and used
   \[
   \nabla^+_i \bar h_N(s, x) = \bar h_N(s,x+N^{-1}e_i) - \bar h_N(s,x).
   \]
Assuming that $\sigma_D(u)$ is a smooth function of $u$ and that $\bar h_N $ converges to $h$ we obtain
   \begin{multline*}
 \varphi_N(t) - \varphi_N(0) \\  \approx {N^{1-d}} \int_0^t  \sum_{\beta\in \mathbb{T}^d_N }  (2d)^{-1} \left[e^{-KN^{-1}\text{div}\left(\sigma_D(\nabla  h(s,\cdot)\right)}\right]_{\frac{\beta}{N}}
\Delta  v \left(N^{-1}{\beta}\right) ds .
   \end{multline*}
For large $N,$ the term on the right is approximated by
\begin{equation*}
-{N^{-d}} \int_0^t  \sum_{\beta\in \mathbb{T}^d_N }  (2d)^{-1}\, K \text{div}\left( \sigma_D (\nabla  h(s,\cdot)\right)_{ |_\frac{\beta}{N}}
\Delta  v \left(N^{-1}{\beta}\right) ds,
\end{equation*}
where we have appealed to the periodic boundary conditions of $v.$

Therefore, in the limit as $N\rightarrow \infty$ we have that
\begin{equation*}
\varphi_N(t) - \varphi_N(0) \approx  - \int_0^t  \int (2d)^{-1}\,K \text{div}\left(\bar \sigma(\nabla  h(s,\cdot)\right)_{ |_x} \Delta  v(x) \, dx\, ds
\end{equation*}
or after integrating by parts on the right hand side and differentiating in time,
\[
\int \left(\partial_t h(t,x) \right) v(x) dx = \int  -(2d)^{-1}\, K\Delta \left[\text{div}\left(\sigma_D(\nabla  h(s,\cdot)\right)\right](x) v(x) dx.
\]
Since this augment can be applied for any test functions, $v,$ we arrive at
\[
\partial_t h = -(2d)^{-1}\, K\Delta \left[\text{div}\left(\sigma_D(\nabla  h\right)\right].
\]

\subsection{The rough scaling regime}
\label{s:roughkrug}

In this section we make the assumption that for some $p>1$ the limit
\begin{equation*}
V^{\infty} (x) = \lim_{\kappa \rightarrow \infty} \kappa^{-p}V(\kappa x)
\end{equation*}
exists and is a smooth function of $x\in\mathbb{R}^d.$ 
In the argument below we will need to characterize the limit of $\kappa^{1-p}\sigma_D(\kappa u)$ for very large $\kappa.$
To that end recall that for any $\kappa>0,$ $\sigma_D$ satisfies
 \begin{equation*}
\kappa u  =  \frac{\sum_{z\in \mathbb{Z}^d} z\, e^{-K \sum_{i\leq d}V(z_i) +   K\sigma_D(\kappa u)^\text{\tiny T} z}}
 {\sum_{z\in \mathbb{Z}^d} e^{-K\sum_{i\leq d} V(z_i) + K\sigma_D(\kappa u)^\text{\tiny T} z}}.
\end{equation*}
Defining $\sigma_\kappa(u) = \kappa^{p-1} \sigma_D(\kappa u),$ this expression can be rewritten as
\begin{align*}
u & =  \frac{\sum_{z\in \mathbb{Z}^d} \kappa^{-1} z\, e^{-K \kappa^p \sum_{i\leq d}V(\kappa^{-1} z_i) +   K\kappa^p \sigma_\kappa(u)^\text{\tiny T} (\kappa^{-1}z)}}
{\sum_{z\in \mathbb{Z}^d} e^{-K\kappa^p \sum_{i\leq d} V(\kappa^{-1} z_i) + K\kappa^p \sigma_\kappa(u)^\text{\tiny T} (\kappa^{-1}z)}}\\
& \approx  \frac{\int  w\, e^{-K \kappa^p \sum_{i\leq d}V( w_i) +   K\kappa^p \sigma_\kappa(u)^\text{\tiny T} w}dw}
{\int  e^{-K\kappa^p \sum_{i\leq d} V(w_i) + K\kappa^p \sigma_\kappa(u)^\text{\tiny T} w}dw}.
 \end{align*}
When $\kappa$ is large the expression on the right converges to the value of $w$ that minimizes
$ \sum_{i\leq d}V( w_i) +   \sigma_\kappa(u)^\text{\tiny T} w.$  In order for this minimum to be equal attained at $u$ we should have
that $\sigma_\kappa(u)$ converges to $\nabla V(u).$
Thus we define
\[
\bar \sigma(u) = \lim_{\kappa\rightarrow \infty} \kappa^{1-p} \sigma_D( \kappa u) = \nabla V(u).
\]

Now set
\[
q = \frac{p}{p-1}
\]
and, as in the previous subsection, consider sums of $h_N$ against sampled values of a smooth periodic function $v,$
\[
\varphi_N(t) = N^{-(1+d)} \sum_{\alpha\in \mathbb{T}^d_N} h_N(N^4 t,\alpha) v(N^{-1}\alpha).
\]
Appealing to the smoothness of $v$ we have that
\[
\varphi_N(t) \approx \int \bar h_N(t,x) v(x) dx
\]
where in this subsection the overbar represents the projection defined in \eqref{roughscaling}.

By exactly the same arguments as in the previous section we arrive at the formula
   \begin{multline*}
\varphi_N(t) - \varphi_N(0) \approx {N^{-d}} \int_0^t  \sum_{\beta\in \mathbb{T}^d_N }  (2d)^{-1}  \Delta  v \left(N^{-1}{\beta}\right) \\\times e^{-K \sum_{i\leq d}\sigma_{D,i}(N^q\nabla^+\bar h_N(s, N^{-1}\beta))- \sigma_{D,i}(N^q  \nabla^+ \bar h_N(s, N^{-1}(\beta-e_i)))} ds , 
   \end{multline*}
   where again we have used
   \[
   \nabla^+_i \bar h_N(s, x) = \bar h_N(s,x+N^{-1}e_i) - \bar h_N(s,x).
   \]
   Writing $N^q$ as $N^{q-1} N$ (note that $(q-1)(p-1)=1$) and assuming that 
   $ N \nabla^+\bar h_N(s, N^{-1}\beta) $ and 
   $ N  \nabla^+ \bar h_N(s, N^{-1}(\beta-e_i)) $
   are approximations of the 
   derivative of a smooth function we can use the approximation 
   \[
   \bar \sigma (u) \approx N^{(q-1)(1-p)} \sigma_D(N^{q-1} u) = N^{-1} \sigma_D(N^{q-1} u)
   \]
    to conclude that
      \begin{align*}
&  \varphi_N(t) - \varphi_N(0) 
\approx {N^{-d}} \int_0^t  \sum_{\beta\in \mathbb{T}^d_N }  (2d)^{-1} \Delta  v \left(N^{-1}{\beta}\right) \\
& \hspace{2.0cm} \times e^{- K N \sum_{i\leq d}( \bar\sigma_i(N\nabla^+\bar h_N(s, N^{-1}\beta))- \bar\sigma_i(N  \nabla^+ \bar h_N(s, N^{-1}(\beta-e_i))))}ds.
   \end{align*}
   Our assumption that $\bar\sigma$ is a smooth function then suggests that
     \begin{align*}
& \varphi_N(t) - \varphi_N(0) \approx \\
& \hspace{1.5cm} {N^{-d}} \int_0^t  \sum_{\beta\in \mathbb{T}^d_N }  (2d)^{-1}
 \left[ e^{-K \text{div}\left(\bar \sigma(\nabla h(s,\cdot)\right)}\right]_\frac{\beta}{N}\Delta  v \left(N^{-1}{\beta}\right) ds
   \end{align*}
   where $h$ is the limit of $\bar h_N.$
   In the large $N$ limit we have that
  \begin{equation*}
 \varphi_N(t) - \varphi_N(0)   =  \int_0^t \int (2d)^{-1}\,\left[e^{- K \text{div}\left(\bar \sigma(\nabla h(s,\cdot)\right)}\right]_x \Delta  v(x)\, dx\,ds
  \end{equation*}
  or, after integrating by parts and differentiating,
  \begin{equation*}
   \partial_t h  = (2d)^{-1}\Delta\left[ e^{- K \text{div}\left(\bar\sigma^\infty(\nabla h)\right)} \right].
\end{equation*}

\end{document}